\documentclass[useAMS,usenatbib]{mn2e}
\pdfoutput=1
\usepackage{times}
\usepackage{epsfig}
\usepackage{amssymb,amsmath}
\usepackage{graphicx}
\usepackage{enumerate}
\usepackage{amsmath}
\usepackage{longtable,lscape}
\usepackage{rotating}
\usepackage{multirow,multicol}
\usepackage[breaklinks]{hyperref}
\usepackage{natbib}
\usepackage{color}
\usepackage{amsmath}
\usepackage{color,soul}
\usepackage{comment}

\title[Characteristics of red submillimetre galaxies]{H-ATLAS/GAMA: 
The nature and characteristics of optically red galaxies detected at submillimetre wavelengths}
\author[Dariush et al. ]
{
A. Dariush$^{1}$\thanks{AD: adariush@ast.cam.ac.uk},
S. Dib$^{2}$\thanks{SD; sdib@nbi.dk}, 
S. Hony$^{3}$\thanks{SH; sacha.hony@free.fr}, 
D. J. B. Smith$^{4}$, S. Zhukovska$^{5}$, L. Dunne$^{9,25}$, 
\newauthor
S. Eales$^{8}$, E. Andrae$^{28}$, M. Baes$^{14}$, I. Baldry$^{27}$, A. Bauer$^{6}$, J. Bland-Hawthorn$^{23}$, 
\newauthor
S. Brough$^{10}$,	N. Bourne$^{25}$,   A. Cava$^{18}$, D. Clements$^{7}$,  M. Cluver$^{24}$, A. Cooray$^{17}$, 
\newauthor
G. De Zotti$^{16}$,	S. Driver$^{20,33}$, M. W. Grootes$^{28}$, A. M. Hopkins$^{6}$, R. Hopwood$^{7,22}$,
\newauthor
 S. Kaviraj$^{7, 13}$, L. Kelvin$^{34}$, M. A. Lara-Lopez$^{32}$ , J. Liske$^{26}$, J. Loveday$^{31}$, S. Maddox$^{9}$,
\newauthor
	 B. Madore$^{12}$, M.~J.~Micha{\l}owski$^{15}$, C. Pearson$^{13,19,22}$, C. Popescu$^{29,30}$,	A. Robotham$^{20}$,  
\newauthor
K. Rowlands$^{33}$, M. Seibert$^{12}$, F. Shabani$^{21}$, M. W. L. Smith$^{8}$,  E.N. Taylor$^{11}$, R. Tuffs$^{28}$, 
\newauthor
E. Valiante$^{8}$, J.S. Virdee$^{13,19}$\\ \ \\
$^{1}$Institute of Astronomy, University of Cambridge, Madingley Road, Cambridge, CB3 0HA, UK\\
$^{2}$Niels Bohr Institute \& Centre for Star and Planet Formation, University of Copenhagen, {\O}ster Voldgade 5-7., DK-1350, Copenhagen, Denmark\\
$^{3}$Zentrum f\"{u}r Astronomie der Universit\"{a}t Heidelberg, Institut f\"{u}r Theoretische Astrophysik, Albert-Ueberle-Stra{\ss}e 2, 69120 Heidelberg, Germany\\
$^{4}$Centre for Astrophysics, Science \& Technology Research Institute, University of Hertfordshire, Hatfield, Herts, AL10 9AB, UK\\
$^{5}$Max-Planck-Institut f\"ur Astrophysik, Karl-Schwarzschild-Str. 1, D-85741 Garching, Germany\\
$^{6}$Australian Astronomical Observatory, PO Box 915, North Ryde, NSW 1670, Australia\\
$^{7}$Physics Department, Imperial College London, Prince Consort Road, London, SW7 2AZ, UK\\
$^{8}$School of Physics and Astronomy, Cardiff University, the Parade, Cardiff, CF24 3AA, UK\\
$^{9}$Department of Physics and Astronomy, University of Canterbury, Private Bag 4800, Christchurch, 8140, NZ\\
$^{10}$Australian Astronomical Observatory, PO Box 296, Epping, NSW, 1710, Australia\\
$^{11}$School of Physics, the University of Melbourne, Parkville, VIC 3010, Australia\\
$^{12}$Observatories of the Carnegie Institution for Science, 813 Santa Barbara Street, Pasadena, CA 91101, USA\\
$^{13}$Oxford Astrophysics, Denys Wilkinson Building, University of Oxford, Keble Rd, Oxford OX1 3RH\\
$^{14}$Sterrenkundig Observatorium, Universiteit Gent, Krijgslaan 281S9, 9000 Gent, Belgium\\
$^{15}$SUPA, Institute for Astronomy, University of Edinburgh, Royal Observatory, Edinburgh, EH9 3HJ\\
$^{16}$INAF-Osservatorio Astronomico di Padova, Vicolo Osservatorio 5, I-35122 Padova, Italy, and SISSA, Via Bonomea 265, I-34136 Trieste, Italy\\
$^{17}$Department of Physics and Astronomy, University of California, Irvine, CA 92697 USA\\
$^{18}$Observatoire de Gen{\`e}ve, Universit{\'e} de Gen{\`e}ve, 51 Ch. des Maillettes, 1290 Versoix, Switzerland\\
$^{19}$RAL Space, STFC Rutherford Appleton Laboratory, Harwell Oxford, Didcot, Oxon, OX11 0QX, UK\\
$^{20}$International Centre for Radio Astronomy Research (ICRAR),University of Western Australia, 35 Stirling Highway, Crawley,WA 6009, Australia \\
$^{21}$Finnish Centre for Astronomy with ESO (FINCA), University of Turku, V$\ddot{a}$is$\ddot{a}$l$\ddot{a}$ntie 20, FI-21500 Piikki$\ddot{o}$, Finland\\
$^{22}$Department of Physical Sciences, The Open University, Milton Keynes, MK7 6AA, UK \\
$^{23}$Sydney Institute for Astronomy, School of Physics A28, University of Sydney, NSW 2006, Australia\\
$^{24}$University of the Western Cape, Robert Sobukwe Road, Bellville, 7535, South Africa\\
$^{25}$Institute for Astronomy, University of Edinburgh, Royal Observatory, Blackford Hill, Edinburgh EH9 3HJ, UK\\
$^{26}$European Southern Observatory, Karl-Schwarzschild-Str. 2, 85748 Garching, Germany\\
$^{27}$Astrophysics Research Institute, Liverpool John Moores University, IC2, Liverpool Science Park, 146 Brownlow Hill, Liverpool, L3 5RF\\
$^{28}$Max Planck Institut fuer Kernphysik, Saupfercheckweg 1, 69117 Heidelberg, Germany\\
$^{29}$Jeremiah Horrocks Institute, University of Central Lancashire, Preston, PR1 2HE, UK\\
$^{30}$The Astronomical Institute of the Romanian Academy, Str. Cutitul de Argint 5, Bucharest, Romania\\
$^{31}$Department of Physics and Astronomy, University of Sussex, Brighton, BN1 9RH\\
$^{32}$Instituto de Astronom\'ia, Universidad Nacional Autónoma de M\'exico, A.P. 70-264, 04510 M\'exico, D.F., M\'exico\\
$^{33}$School of Physics \& Astronomy, University of St Andrews, North Haugh, St Andrews, Fife, KY16 9SS, UK\\
$^{34}$Institut f\"{u}r Astro- und Teilchenphysik, Universit\"{a}t Innsbruck, Technikerstra{\ss}e 25, 6020 Innsbruck, Austria
}

\def\add{\textcolor{black}}

\def\revised{}

\begin{document}


\date{}

\pagerange{\pageref{firstpage}--\pageref{lastpage}} \pubyear{2002}

\maketitle

\label{firstpage}


\def	\submm			{submm }
\def	\Submm			{Submm }
\def\sersic     		{~S\'{e}rsic }
\def	\deg2			{~deg$^2$}
\def	\ki2				{~$\chi^2$}
\def	\micron			{$\rm{\mu}$m}
\def	\Msun			{~M$_{\sun}$}
\def	\sfr				{~$\psi$/M$_{\sun}$yr$^{-1}$}
\def	\sfrunit			{~M$_{\odot}$yr$^{-1}$}
\def	\ssfrunit		{~yr$^{-1}$}
\def	\3sigma			{~$>$5$\sigma$}
\def	\rpetro	    		{~$r_{_{\rm Petro}}$}
\def\logMstar   		{$\log (\rm{M}_{*}/\rm{M}_{\sun})$}
\def\logMdust   		{$\log (\rm{M_{\rm D}}/\rm{M_*})$}
\def\Mdust      		{$\rm{M_{\rm D}}/\rm{M_*}$}
\def\nuvr       		{~NUV$-r$}

\def	\fmu				{~f$_\mu$}
\def\mstar   		{$\rm{M}_*/\rm{M}_{\sun}$}
\def\mdust   		{$\rm{M}_{\rm {\rm D}}/\rm{M}_{\sun}$}
\def\mmdust   		{$\rm{M}_{\rm {\rm D}}$}
\def	\ssfr			{~SSFR/yr$^{-1}$}
\def\M*         		{M$_{*}$ }
\def\rg				{{\it red }}
\def	\bg				{{\it blue }}
\def	\pg				{{\it passive }}
\def	\cg				{{\it control }}
\def\path        	{./}

\clearpage
\begin{abstract}

We combine Herschel/SPIRE sub-millimeter (submm) observations with existing multi-wavelength data to investigate the characteristics  of low redshift, optically red galaxies detected in \submm bands. We select a sample  of galaxies in the  redshift range 0.01~$\leq$~z~$\leq$~0.2, having \3sigma ~detections in the SPIRE 250 \micron~ \submm waveband. Sources are then divided into two sub-samples of \rg and \bg galaxies, based on their UV-optical colours. Galaxies in the \rg sample account for $\approx$~4.2 per cent of the total number of sources with stellar masses M$_{*}\gtrsim$~10$^{10}$M$_{\sun}$. Following visual classification of  the \rg galaxies, we find that $\gtrsim$~30 per cent of them are early-type galaxies and $\gtrsim$~40 per cent are  spirals.  The colour of the \rg-spiral galaxies could be the result of their highly inclined orientation and/⁠or a strong contribution of the old stellar population.

It is found that irrespective of their morphological types, \rg and \bg sources occupy environments with more or less similar densities (i.e., the $\Sigma_5$ parameter). From the analysis of the spectral energy distributions (SEDs) of galaxies in our samples based on MAGPHYS, we find that galaxies in the \rg sample (of any morphological type) have dust masses similar to those in the \bg sample (i.e. normal spiral/star-forming systems). However, in comparison to the {\it red}-spirals and in particular \bg systems, {\it red}-ellipticals have lower mean dust-to-stellar mass ratios. {\revised Besides galaxies in the {\it red}-elliptical sample have much lower mean star-formation/specific-star-formation rates in contrast to their counterparts in the \bg sample.}  Our results support a scenario where dust in early-type systems is likely to be of an external origin.

\end{abstract}


\section{Introduction}
\label{sec:intro}

Galaxies display a wide variety of physical and observational properties. It is well known  that the distribution of galaxy optical colours is bimodal, e.g.  blue cloud versus the red sequence \citep{stra,baldry,taylor15}. The bimodality of the galaxy population  exists at least out to $z\simeq$~1 \citep[e.g. ][]{Bell04a,Tanaka05,Cooper06,Cucciati06,Willmer06}. A number of different mechanisms (taking place in different environments) have been proposed for the observed bimodality of the galaxy population, including, but not limited to, galaxy merging (major and minor), galaxy strangulation and harassment, ram-pressure stripping as well as AGN feedback  \citep[e.g. ][]{Mulchaey00,Croton06,Conselice14}. Such mechanisms could regulate the observed optical colours of galaxies by influencing their key physical parameters such as star formation history (SFH), mean age of stellar populations, the amount of dust attenuation, dust geometry and metallicity \citep{bruzual03,burgarella05,ed08,conroy09}. \\ \ \\

Besides, there are substantial differences between galaxy populations in the field and those in clusters and groups. According to \citet{dressler80}, galaxy morphology is a strong function of galaxy density, i.e.  the morphology-density relation, and numerous studies since then have shown the dependence of galaxy properties on the local environment \citep{binggeli87,lewis02,baloghb,ball}. For example, the red population is substantially dominated by early-type galaxies and thus preferentially found in high-galaxy density environments, while blue galaxies are predominantly late-type systems and mostly found in low-galaxy density environments, i.e. the colour-density relation. Moreover, vast majority of galaxies in the blue cloud are actively forming stars while the red sequence consists mainly of passive galaxies with little or no ongoing star formation. There are also  additional contributions to the red cloud from (a) heavily obscured star-forming or edge-on galaxies and (b) galaxies with passive disks, e.g. red spirals showing signs of low-level of star formation, which are known to be considerably redder and more massive than their blue/star-forming counterparts \citep{bergh76,wolf09,m10,luca12a}. It is noteworthy that the morphology-density and color-density relations evolve with redshift \citep[e.g. ][]{Butcher84,Poggianti09,Poggianti10}.

Analyses of the dust attenuation in active/star-forming galaxies suggest that in contrast to passive galaxies, they are heavily affected by dust \citep{driver07,johnson07,wyder,luca,tojeiro09,grootes13}.   It has been shown that the bulk of the dust in late-type galaxies is in the cold phase and as consequence emits at $>$100 \micron , i.e. the far-infrared (FIR) and \submm wavelengths \citep{sodroski97,odenwald98,dunne01,popescu02,popescutuffs02,vlahakis05,dale07,dale12,bendo12}. Such wavelengths are covered by the instruments on board the {\it Herschel Space Observatory} \citep{pil}\footnote{Herschel is an ESA space observatory with science instruments provided by European-led Principal Investigator consortia and with important participation from NASA.}. Thus, the data collected by {\it Herschel} is uniquely suited to probe the dusty component, e.g. its characteristics and origin, in all type of galaxies, in particular early-type galaxies which contain significantly less dust than late-type systems.

The existence of dust in early-type galaxies has been first reported from studying the absorption of stellar light \citep{Bertola78,Ebneter85,Goudfrooij94} and since then several studies have been conducted in order to shed light on the quantitative
dust content of eary-type galaxies\citep{Knapp89,Temi04,Temi07,Leeuw04,Savoy09}. However, \submm data provided by {\it Herschel} have enabled us to study dust properties, e.g. its total luminosity, mass and temperature in early-type galaxies in an unprecedented manner due to a better sensitivity, resolution and/or the long wavelength coverage necessary \citep{Boselli10,Davies10,DeLooze10,Auld12,Smith12b,Alighieri13}.

Among various surveys,  the {\it Herschel} Astrophysical Terahertz Large Area Survey \citep[H-ATLAS;][]{se} is the widest extragalactic survey  undertaken in \submm with {\it Herschel}. The large coverage of H-ATLAS helps to have a better statistical view of the dust content and its characteristic among galaxies spanning a broad range of luminosities, colours and morphologies. Results from \citet{ad11} as part of the H-ATLAS Science Demonstration Phase (SDP) and based on the UV-optical colour classification, show that the majority of sources ($\simeq$~95 per cent) with \submm detections at low redshift (z~$\leq$~0.2), are blue/star-forming galaxies with UV-optical colours \nuvr~$\leq$~4.5. This earlier study suggested that the submm-detected/optically-red galaxies (\nuvr~$>$~4.5), with a contribution of $\lesssim$~5 per cent to the total number of detections, are more likely to be star-forming galaxies and that their red colors are due to obscuration by dust. 

From a stacking analysis at \submm wavelengths, \citet{nb12} performed a large-scale statistical study of the \submm properties of optically selected galaxies (based on the rest-frame color $g-r$) at z~$\lesssim$~0.35, and found that  approximately  20 per cent of low-redshift galaxies  in H-ATLAS are red.

In the mean time, there have been several H-ATLAS studies trying to shed light on the existence  and properties of dust in early-type galaxies. For instance \citet{kr12} used data from the H-ATLAS SDP to study dust properties and star formation histories in a sample of low redshift galaxies (z~$\lesssim$~0.5)  detected at \submm wavelengths. Followed by classification of their sample based on optical morphology, \citet{kr12} found that $\simeq$~4.1 per cent of all detections are early-type systems and that $\simeq$~3.8 per cent (19 out of 496) of spiral galaxies with \submm detections are passive. In another study and by using samples of early-type galaxies at low redshifts (0.013~$\lesssim$~z~$\lesssim$~0.06), \citet{Agius12} found that early-type galaxies with H-ATLAS detections (based on Phase 1 Version 2.0 internal release of the H-ATLAS catalogue), are not only bluer in the UV-optical colours but also are significantly brighter in NUV in comparison to their H-ATLAS none-detected counterparts.


The aim of this work is to examine in more detail the nature of \submm detected red galaxies using the data of H-ATLAS. The main difference between this work and those conducted by \citet{kr12} and \citet{Agius12} is that all sources in our sample are detected in H-ATLAS and classified  by means of the UV-optical colour index. Our main objectives are: to segregate intrinsically red galaxies from heavily obscured star forming galaxies, and subsequently discuss the origin and the role of the dust in passive systems. The main improvements compared to our previous study come from: 
\begin{itemize}
 \item a larger area coverage (by a factor of $\sim 10$) and therefore a better statistics
 \item the inclusion of complimentary wavelengths in the mid-infrared (MIR) bands
 \item the extraction of various physical parameters from multi-wavelengths observations of sources by means of the SED fitting.
\end{itemize}
The paper is organized as following: In Section \ref{sec:data},   we present the data from H-ATLAS phase 1 and select a sample of low redshift galaxies,  all detected with {\it Herschel} in the SPIRE 250\micron ~\submm band.  In Section \ref{sec:Analysis}, we select sub-samples of optically blue and red galaxies and analyze their physical characteristics such as star formation activities and dust properties as inferred from fitting their spectral energy distributions.  Our main finding and conclusion are given in Section \ref{sec:conclusion}. Throughout the paper, we assume a concordance CDM cosmology with $H_0$ = 70 km s$^{-1}$ Mpc$^{-1}$, $\Omega_m$ = 0.3 and $\Omega_{\Lambda}$ = 0.7.


\section{Data}
\label{sec:data}

We use data from the H-ATLAS Phase 1 Version 3.0 internal release which contains the IDs of \3sigma ~ SPIRE detections at 250 \micron~ and is reduced in a similar way to the SDP data, as described by  \citet{ibar10},~\citet{Pascale11},~\citet{Rigby11} and \citet{Smith11}. The Phase 1 ID catalogues have been produced in a similar way to \citet{Smith11} and will be presented in \citet[][; in prep]{Bourne15}.

 Initially observed time-line data from SPIRE and PACS instruments  were processed by using the Herschel Interactive Processing Environment (HIPE) based on a custom reduction scripts. High-pass filtering was then applied  to the data time-lines in order to correct the thermal drift in bolometer arrays. Cross-scan time-line observations were projected by using the naive map-making method of HIPE. For point like sources, catalogue of \3sigma ~ \submm fluxes were produced from the 250 \micron~  PSF filtered map, using the MADX algorithm (Maddox et al. in prep), as described in \citet{Rigby11}.  For extended sources, larger apertures were chosen such that they match the extent of the source \submm~ emission. For each 250 \micron~ source, corresponding 350 and 500 \micron~ flux densities were estimated by using the 350 and 500 \micron~ maps (noise- weighted/beam-convolved) at the source position extracted from the 250 \micron~ map. Finally 100 and 160 \micron~ aperture flux densities were measured following matching each 250 \micron~ source to the nearest PACS sources within a radius of 10 arcsec. A likelihood-ratio analysis \citep{Sutherland92} was performed by \citet{Bourne15} to match 250 \micron~ sources to the SDSS DR7 \citep{Abazajian09} sources brighter than r = 22.4 mag within a 10 arcsec radius. The probability that an optical source is associated with the \submm source has been used to define the reliability of an association. According to \citet{Bourne15}, objects with reliability~$\geq$ 0.8 are considered to be true matches to \submm sources.

The H-ATLAS fields are along the celestial equator centred at RA of 9h(G09), 12h(G12) and 14.5h(G15). 
144 \deg2 ~out of the 161 \deg2 ~covered by H-ATLAS overlap with the Galaxy and Mass Assembly (GAMA I) survey \citep{sd09,sd11}. The GAMA survey re-processes and  combines optical data from the Sloan Digital Sky Survey \citep[SDSS DR6;][]{am}, NIR data from the UKIRT Infrared Deep Sky Survey (UKIDSS) Large Area Survey \citep[LAS DR4;][]{law}, and UV from the Galaxy Evolution Explorer \citep[GALEX;][]{morris}.  The pre-processing of the GAMA, SDSS and UKIDSS archive data is descibed in detail in \citet{Hill10}. For all galaxies with $r\leq$~19.4 mag in G09 and G15 as well as $r\leq$~19.8 mag in G12, redshifts have been measured using the Anglo Australian Telescope and for brighter galaxies, redshift estimates are taken from other existing redshift surveys such as SDSS, the 2dF Galaxy Redshift Survey (2dFGRS) and the Millennium Galaxy Catalogue \citep[MGC;][]{Liske03,Driver05}. Furthermore, the GAMA-{\it WISE} ~\citep[the Wide-field Infrared Survey Explorer;][]{wright10} catalogue   adds coverage in four MIR bands at 3.4\micron,  4.6\micron, 12\micron~ and 22\micron  ~\citep{cluver14}. 

\begin{figure*}
\hspace{-5.0em}
\includegraphics[scale=0.4]{\path 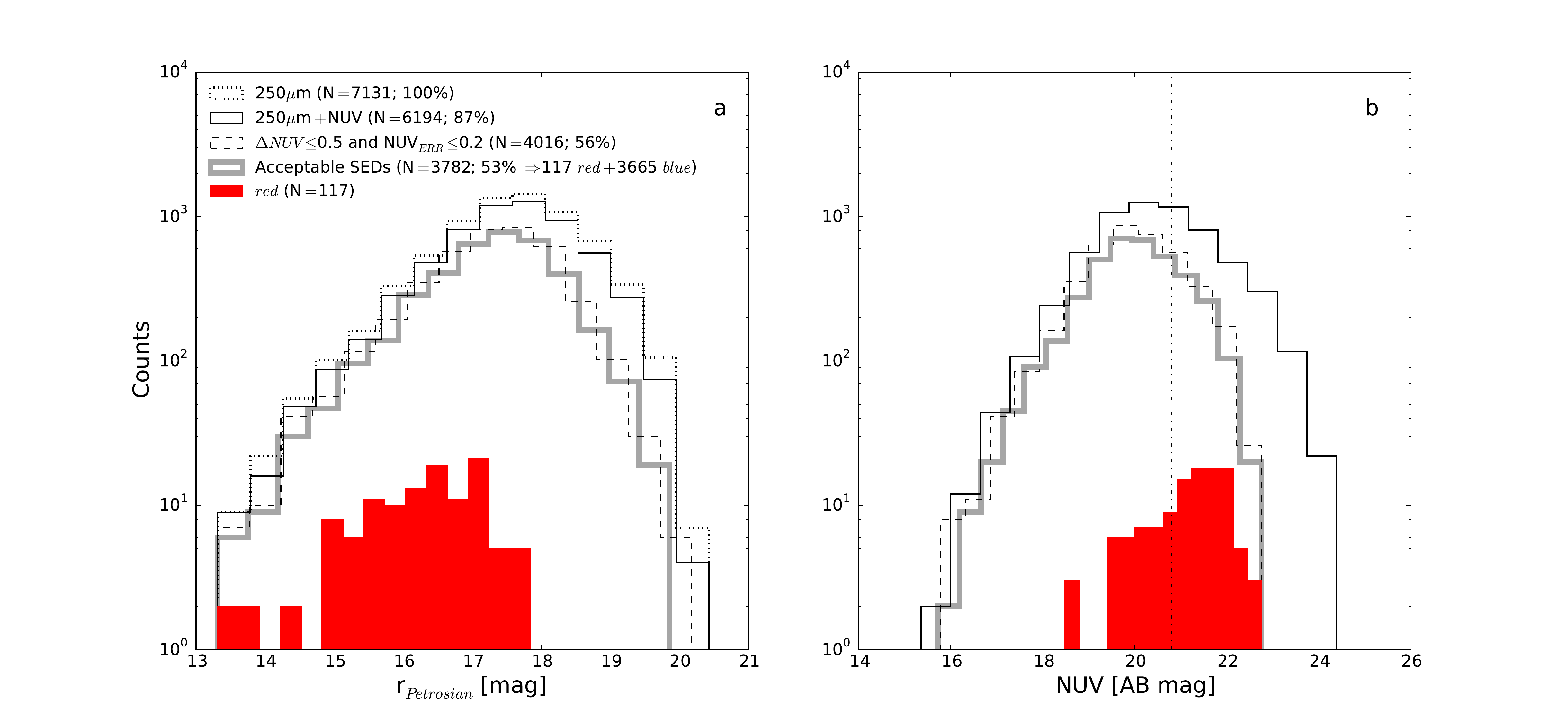}
\caption{Distributions of the SDSS $r$-band (panel a) and NUV (panel b) magnitudes for all galaxies as well as those qualified to be included for the subsequent data analysis. 
'Dotted-line' represents all galaxies detected in 250 \micron~while the 'black-solid-line' shows those observed in NUV with a subset of them (dashed-line) having NUV errors $\leq$0.2 and $\Delta$NUV$\leq$0.5 (i.e. the absolute difference between the GAMA and {\it GALEX} NUV flux measurements). Finally the 'gray-thick-line' represents sources with good quality SED fits as described in Sec.~\ref{sec:SED}. Sources were also divided into two categories of \rg (filled histogram) or \bg based on their UV-optical \nuvr~ colours as discussed in Sec.~\ref{sec:red-selection}. The vertical 'dashed-dotted' line in panel b shows the {\it GALEX} AIS (All-Sky Imaging Survey) NUV depth which is around $\approx$~20.8 magnitude.}
\label{fig:petrosian-r}
\end{figure*}

In summary, we have at our disposal UV, optical and MIR data as well as redshift estimates for the \submm  galaxies within the H-ATLAS/GAMA-overlapping area where all \submm selected sources in our sample satisfy the following criteria:

\begin{itemize}

\item They all have \3sigma ~ \submm detected at SPIRE 250 \micron. 

\item They fall within the redshift range $0.01\leq z \leq 0.2$.  We only select objects with a sufficiently reliable spectroscopic determination \citep[i.e. $n_{Q}\geq$3; ][]{sd11}.

\item All \submm galaxies have a reliability parameter (\texttt{reliability}~$\geq$ 0.8) of being associated with an optical counterpart in the SDSS r-band catalogue, for 
which multi-wavelength photometry is available. As such, in addition to the 250 \micron~ emission, all sources (7131 objects) have corresponding fluxes (all corrected for Galactic extinction) via aperture matched photometry in other bands ranging from UV to MIR. 

\item  Since a crucial aspect of our selection of red galaxies is based on the UV-optical (\nuvr)  colour,  we remove from our sample those galaxies for which their NUV fluxes as estimated in GAMA, differ by more than $>$0.5 magnitude from those retrieved through {\it GALEX} GR6 Data Release based on the All-Sky Imaging survey (AIS) data products (NUV depth $\sim$~20.8 mag). In addition, all selected sources have NUV magnitude errors, as provided by {\it GALEX}-GR6, which are $\leq$ 0.2 mag. This guarantees that all sources in our sample have enough signal-to-noise ratio in UV. The above constraints on UV fluxes, reduces our sample to 4016 sources.

\item Finally since the physical parameters inferred for each galaxy are based on SED fitting techniques, an extra criterion has been applied in order to exclude sources (234 in total) with 
poor quality SED fits (see Sec.~\ref{sec:SED}).

\end{itemize} 
After applying these selection criteria, we find $3782$ galaxies with detections in at least NUV + {\it u, g, r, i, z} and 250 \micron ~ bands. Distributions of the SDSS $r$-band and NUV magnitudes for all galaxies as well as those qualified to be included for the subsequent data analysis are shown in panels of Fig.\ref{fig:petrosian-r}. According to the first panel, approximately $\approx$~13 per cent of the initial \submm sources were excluded following the requirement of a UV detection  for inclusion in the sample. But that does not seems to exclude systematically any particular type of sources as a  Kolmogorov-Smirnov test (KS test) suggests  a $\gtrsim$~70.0 per cent probability that the distribution of sources detected at 250 \micron~ is  similar to the one being observed simultaneously in the 250 \micron $+$ NUV bands. However by limiting errors in the NUV band to $\lesssim$~0.2 mag, more sources ($\approx$ 31 per cent) are excluded in particular faint objects in the NUV band.

A subset of sources have also detections in {\it GALEX} FUV, PACS (100~\micron, 160~\micron) and  SPIRE (350~\micron, 500~\micron) \submm bands. {\it WISE} data are available and recently have been cross-matched, with extended sources from {\it WISE} accounted for correctly, for all GAMA fields. Yet at the time of analysing  galaxy SEDs in this work, {\it WISE} data were only available for the G12 and G15 fields. Thus, out of the 3782 sources, 2622 ($\approx$ 70 per cent) have also aperture-matched {\it WISE}-MIR data. 

\section{analysis}
\label{sec:Analysis}

\subsection{Selection of intrinsically red objects}
\label{sec:red-selection}

Though the vast majority of galaxies at low redshift with \submm detection are star-forming and optically blue, a small fraction of them are red in optical bands (e.g. $u-r$, $g-r$). We separate blue and red galaxies in the sample using the UV-optical index. This is more robust than optical colour indices as it is more  sensitive to recent star-formation activity \citep[e.g.,][]{kaviraj07}. \citet{ad11} separate red and blue galaxies in the H-ATLAS sample at \nuvr~$=$~4.5, estimated through fitting a double Gaussian to the \nuvr~ colour distribution of galaxies, with redshifts $0.01\leq z \leq 0.2$ (i.e. similar to the present work), in the H-ATLAS SDP data. Hence  any source with NUV$-r\geq 4.5$ mag is considered as {\it red}, while {\it blue} objects are those with NUV$-r< 4.5$ mag. As Fig.\ref{fig:petrosian-r}a shows, the majority of the \rg galaxies in our sample have apparent $r$-band magnitudes $\lesssim$~17.5 mag and NUV magnitudes $\gtrsim$~19.0 mag.

\subsubsection{Contamination by radio AGN}
\label{sec:removing-AGN}

In order to ensure that none of the \submm emission has been contaminated by synchrotron emission from radio-jets hosted by active galactic nuclei (AGNs), 
we find and exclude radio AGN as follows. We cross matched the SDSS position of our sources with those from the full, unfiltered radio-source catalogue of \citet{virdee12}. The radio catalogue consists of all sources detected in the H-ATLAS Phase 1 field by the NRAO VLA Sky Survey \citep[NVSS;][]{condon98} and, as such, contains 7823 sources. The outcome is 117 matches having separations of $<1.0$~arcsec. In order to determine whether the radio emission was consistent with the presence of a radio-loud AGN, we calculated $q_{250}$, defined as: 

\begin{equation}
q_{250}=\log_{10} (\frac{\rm{S}_{250}}{\rm{S}_{1.4}}),
\end{equation}

where $\rm{S}_{250}$ and $\rm{S}_{1.4}$ are fluxes at 250 \micron~ and 1.4-GHz for all matched sources respectively. If $q_{250}<1.4$ then part of the radio-emission 
is due to AGN activity \citep{jarvis10}. Conservatively, we exclude any source which satisfies this criterion in order to ensure none of the submm emission may be contaminated by radio AGN activity. Out of 117 sources with radio counterparts, only 13 sources (1 \rg and 12 \bg galaxies) have $q_{250}<1.4$ and are thus excluded from the subsequent analysis.

\subsubsection{Morphology of the \rg galaxies}
\label{sec:red-morphology}

 The SDSS postage-stamp images of all \rg sources together with their SEDs (inferred as described in Sec.\ref{sec:SED}) are presented in Appendix.\ref{appendix:gallery}. 
 
\begin{figure}
\hspace{-5.0em}
\includegraphics[scale=0.7]{\path 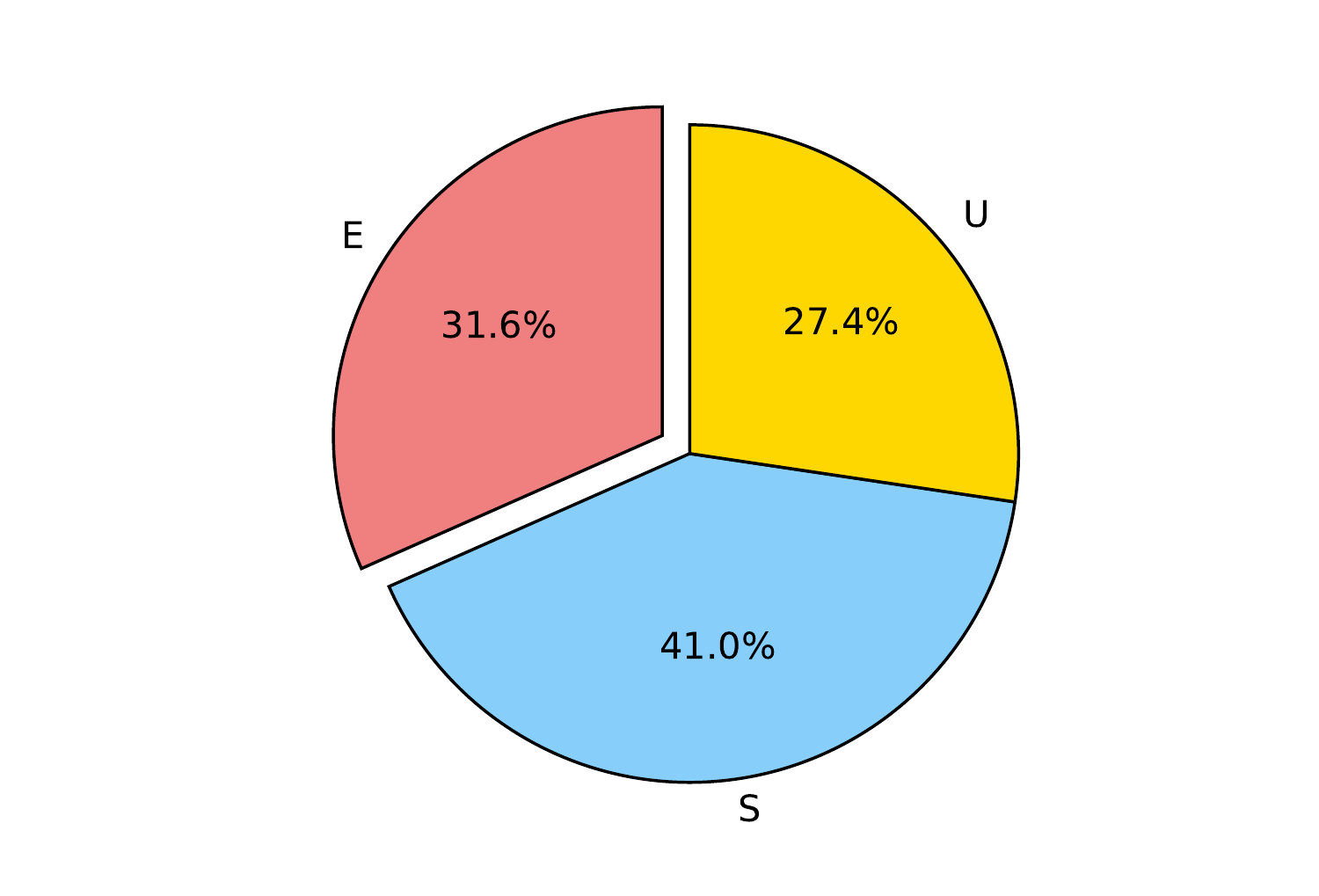}
\caption{Percentage of each morphological type in the sample of 117 \rg galaxies 
(see \S \ref{sec:red-morphology}). Labels represent elliptical (E), spiral (S)
 and undefined (U) galaxies.}
\label{fig:red-morphology-pie-chart}
\end{figure}

The morphology of all 117 galaxies were examined from their SDSS $r$-band images, following independent visual inspection by three team members. Galaxies were classified into three categories of elliptical (E), spiral (S) and uncertain (U). The number of sources in each morphological type is 37, 48 and 32 for the E, S and U galaxies respectively (see Fig.\ref{fig:red-morphology-pie-chart}).  Many of sources classified as U are too small in the SDSS images to judge their morphology and can be of any type, i.e. spiral, elliptical or merging galaxies.

In order to test the validity of this morphological classification, we compared our classification to an independent morphological classification based on the \sersic index $n$ which we obtained from the SDSS DR7 galaxy catalogue  \citet{Simard11}. Different studies have adopted different thresholds of the\sersic index above/below which a galaxy is considered as early/late type. For instance, \citet{Ravindranath04} adopts $n$=2.0 to divide their sample into early and late types though \sersic indices of $n>$2.5 have been also used to describe early-type sources \citep[e.g.,][]{Bell04,Barden05}.

\begin{figure*}
\hbox{\hspace{-15ex}\includegraphics[scale=0.38]{\path 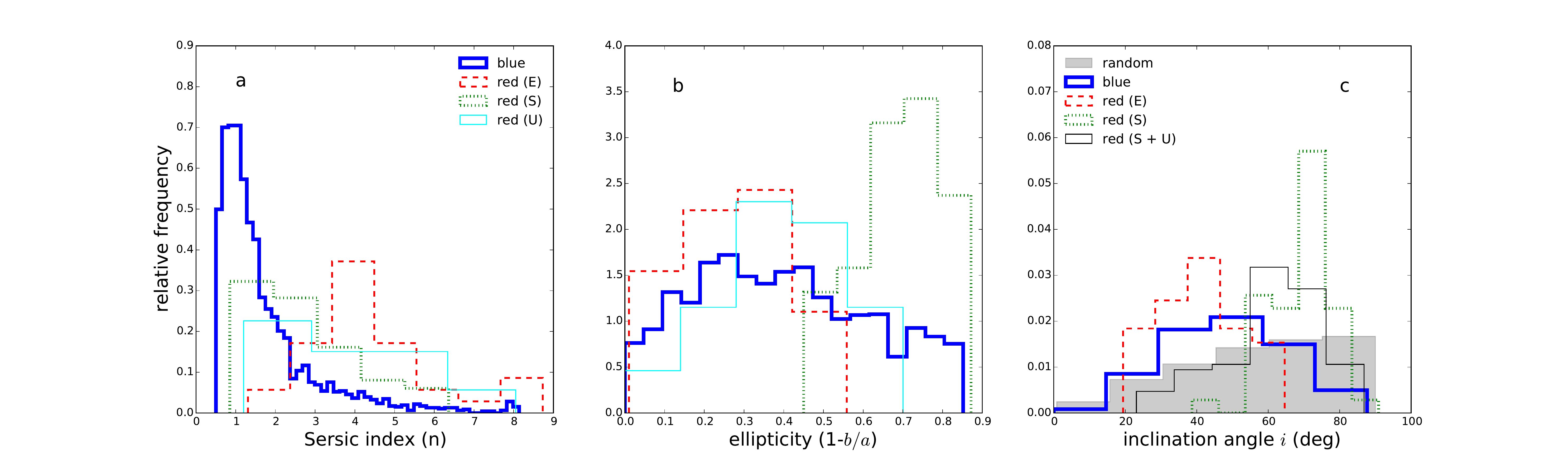}}
\caption{Distributions of morphology related 
 parameters  in all \bg (thick solid line)
 and \rg sources. E (red dashed line), S (green dotted line) and U (cyan line) 
labels represent the morphology of individual \rg source as explained in Sec.\ref{sec:red-morphology}. 
Each histogram is normalized by its integral. Panels represent distributions of galaxy 
(a) \sersic index, (b) ellipticity and (c) inclination angle. In addition, the 'black dotted line' and 'gray filled histogram' in panel c represent the distribution of {\it red}-S+R galaxies and random distribution of inclination angles respectively.}
\label{fig:red-morphology}
\end{figure*}

Fig.~\ref{fig:red-morphology} (panel a) displays the distributions of \sersic indices for all galaxies in our sample, i.e. the \bg sample as well as the morphologically classified \rg galaxies\footnote{We perform a Kolmogorov-Smirnov test (KS test), associated to different estimated parameters, for each pair of galaxy types. The results ($p$-values) are reported in the KS-test Table.~\ref{table:kstest}.}. From this figure, it is clear that the distribution of \sersic indices for the \rg-E sample peaks around $\approx 4$. This is larger than those estimated for the S galaxies (either \bg or \rg). The \sersic index distribution of the \rg-U galaxies lies somewhat between those of the S and E samples.

An inspection of the ellipticity parameter \footnote{The ellipticity for each galaxy has been estimated as ($e=1-b/a$) where $a$ and $b$ are the galaxy's semi-major and semi-minor axes as measured in the SDSS. } of all galaxies in the sample (Fig.~\ref{fig:red-morphology}, panel b) reveals that,  not surprisingly,  in \rg sources of type S,   $e \gtrsim$ 0.5 whereas in  \rg galaxies of type E,  $e \lesssim$ 0.5. In fact the  disk structure is extremely pronounced in  highly inclined spiral galaxies and therefore the majority of galaxies in the S category are those having larger ellipticities. This is better shown in Fig.~\ref{fig:red-morphology}c where histograms of galaxy inclination angles ($i$) for \bg, \rg-S, \rg-E as well as \rg S+U samples are plotted. Inclinations  are determined from the relation

\begin{equation}
\cos^{2} i = [(b/a)^2 - p^2](1 - p^2)^{-1}
\label{Eq: incl_angle}
\end{equation} 
in which $p$ is the ratio of the smallest to the largest axis of an oblate spheroid of rotation. 
We assume $p$ = 0.20 which is an appropriate value to use for the intrinsic flattening of the distribution
of the light of galactic spheroids \citep[e.g.;][]{vandenbergh88}. 

Unlike \bg and \rg-⁠E galaxies, the majority of \rg-⁠S galaxies are
highly inclined. Note that, even in the combined \rg-⁠U $+$ \rg-⁠S sample,
there is still and excess of galaxies with relatively large inclination angles in comparison to the \bg and \rg-E samples.


To illustrate this, we show in Fig.~\ref{fig:red-morphology}c the distribution of inclination angles as expected from a random sampling. The observed difference between the distribution of \rg-(S$+$U) galaxies in comparison to a sample of simulated inclinations, suggests that  the fraction of highly inclined systems in \rg-(S$+$U) sample is more than one would expect for a random distribution. This shows that  the inclination angle play a non-negligible role in the observed red colour of \rg-S systems.

The main conclusion is that the \rg-⁠E sample consists of intrinsically red objects while the \rg-⁠S sample contains galaxies where inclination could be a dominant factor in determining the observed red optical colours. Although these inclined sources are not the main interest of this paper, we do discuss some of their ensemble properties in Sec.~\ref{subsub:sfr-red-s}.

\subsection{Environmental density of \rg galaxies}
\label{sec:density-sigma5}

In order to explore the environmental density of \rg galaxies and see if it plays an important role in shaping their observed properties, we compute the projected surface density around each galaxy. This is  based on counting the number of nearest neighbours, i.e. the density within the distance to the Nth nearest neighbour.  Hence, the surface density to the fifth nearest neighbour is calculated as:

\begin{equation}
\Sigma_{5} ({\rm Mpc^{-2}})=\frac{5}{\pi d^2_{5}},
\label{eq:density-sigma5}
\end{equation}
where $d_5$ is the projected co-moving distance to the fifth nearest neighbour within a volume-limited density-defining population and relative velocity $\pm$1000~km~s$^{-1}$  \citep{whb,sb}. The density-defining population (DDP) are galaxies brighter than $M_r\leq -20.0$. Densities are calculated for galaxies with \rpetro$\leq$ 19.4 (where \rpetro ~ is the r-band Petrosian magnitude), $0.01\leq z \leq 0.18$ and with reliable redshifts \citep[$n_{Q} \geq 3$;][]{sd11}. Although Eq.~\ref{eq:density-sigma5} is a 2D estimate, the redshift information of each galaxy is used to remove the background and foreground sources.

\begin{figure}
\hspace{-3.0em}
\includegraphics[scale=0.45]{\path 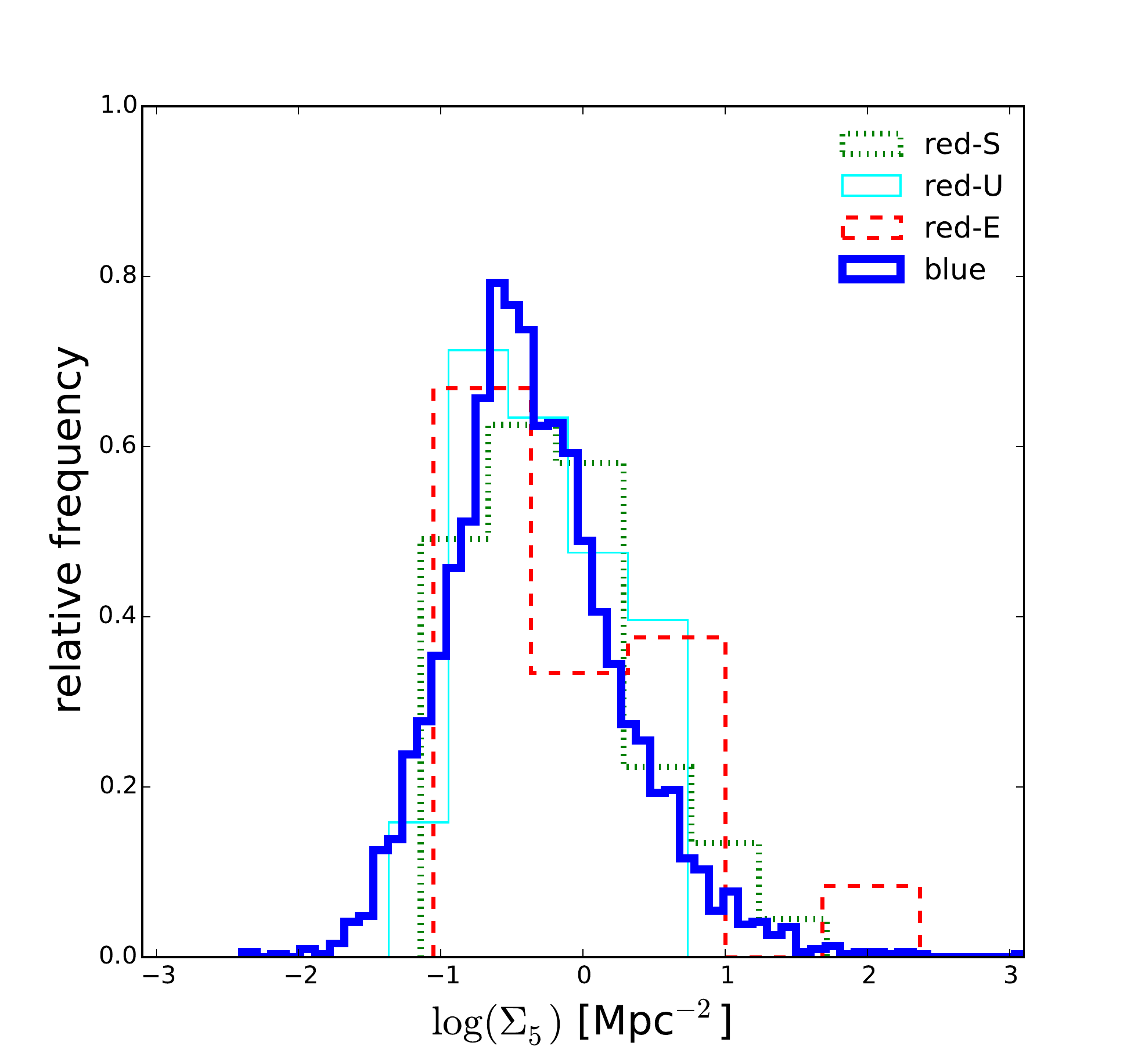}
\caption{Distributions of the projected surface density 
$\Sigma_{5}$ estimated according to Eq.\ref{eq:density-sigma5} in \bg (thick solid line)
 and \rg sources. E (red dashed line), S (green dotted line) and U (cyan line) 
labels represent the morphology of individual \rg source. 
Each histogram is normalized by its integral.}
\label{fig:density-sigma5}
\end{figure}

Fig.~\ref{fig:density-sigma5} displays histograms of the projected densities  for \bg and \rg galaxies within the redshift range of $0.01\leq z \leq 0.18$ and for all systems having $M_r\leq$~$-20.0$. This decreases the overall number of \rg galaxies  by $\approx$~4.2 per cent (out of 117 \rg galaxies, two have $z >0.18$ and three have $M_r>$~$-20.0$). Although the highest observed density (1.5 $\lesssim \log(\Sigma_{5}) \lesssim$ 2.5) is populated by a small fraction of the \rg-E type systems which indeed are relatively massive galaxies, there is no significant difference between the distribution for the \rg sources in any morphological type with respect to the one corresponding to the \bg sample. This indicates that all galaxies, irrespective of their morphologies,  reside in environments with similar densities. It is worth mentioning however that within the redshift range considered here, the survey area does not contain very dense, cluster-like, environments.

\subsection{UV-to-Submm SED fitting}
\label{sec:SED}

We derive the basic properties of galaxies by fitting their SEDs which makes use of the data (\S.~\ref{sec:data}) going from the NUV up to all available Herschel bands. The SED of each galaxy is fitted using  MAGPHYS \citep[Multi-wavelength Analysis of Galaxy Physical Properties;][]{ed08}. MAGPHYS infers the galactic properties by matching the observed SED with a large library of calculated SEDs. These templates are constructed by considering the spectral evolution of stellar populations that are born with a \citet{Chabrier03} initial mass function (IMF) in combination with infrared dust spectral libraries as described in \citet{ed08}. The model assumes that the energy from UV-optical radiation emitted by the stellar populations is absorbed by dust and  re-radiated in the FIR. It uses also the two-component dust model of \citet{Charlot2000} in order to account for the attenuation of starlight by dust. The model also accounts for the enhanced attenuation of stellar radiation for stars located in star forming regions in comparison to older stars found elsewhere within the galaxy .

As the MAGPHYS analysis is based on AB magnitudes, all available photometry (aperture matched) has been converted to the AB magnitude system before estimating their associated fluxes in units of Jansky (Jy). Additional errors have been added to non-\submm  fluxes before running MAGPHYS to account for the total flux measurements and calibrations between the different surveys. These include adding 10 per cent of the flux values in quadrature for all optical-NIR bands and 20 per cent for the UV bands. For each output parameter, MAGPHYS produces a probability density function (PDF), in addition to the median value of each PDF. The 16th and 84th percentiles of the PDF have been considered as a measure of the uncertainty.

\citet{Smith12a} showed that it is insufficient to identify bad SED fits based on a simple $\chi^2$ threshold, instead deriving a threshold which depends on the number of bands of photometry available, above which there is $< 1$\,per cent chance that the photometry is consistent with the MAGPHYS model. Sources exceeding this varying threshold are identified as bad fits, and excluded from the subsequent analysis. We use the H-ATLAS SED fits over the entire phase 1 area, derived using the same method as in \citet{Smith12a}, with updated PACS coverage and including data from {\it WISE}.

\begin{figure*}
\hspace{-0.5em}
\includegraphics[scale=0.25]{\path 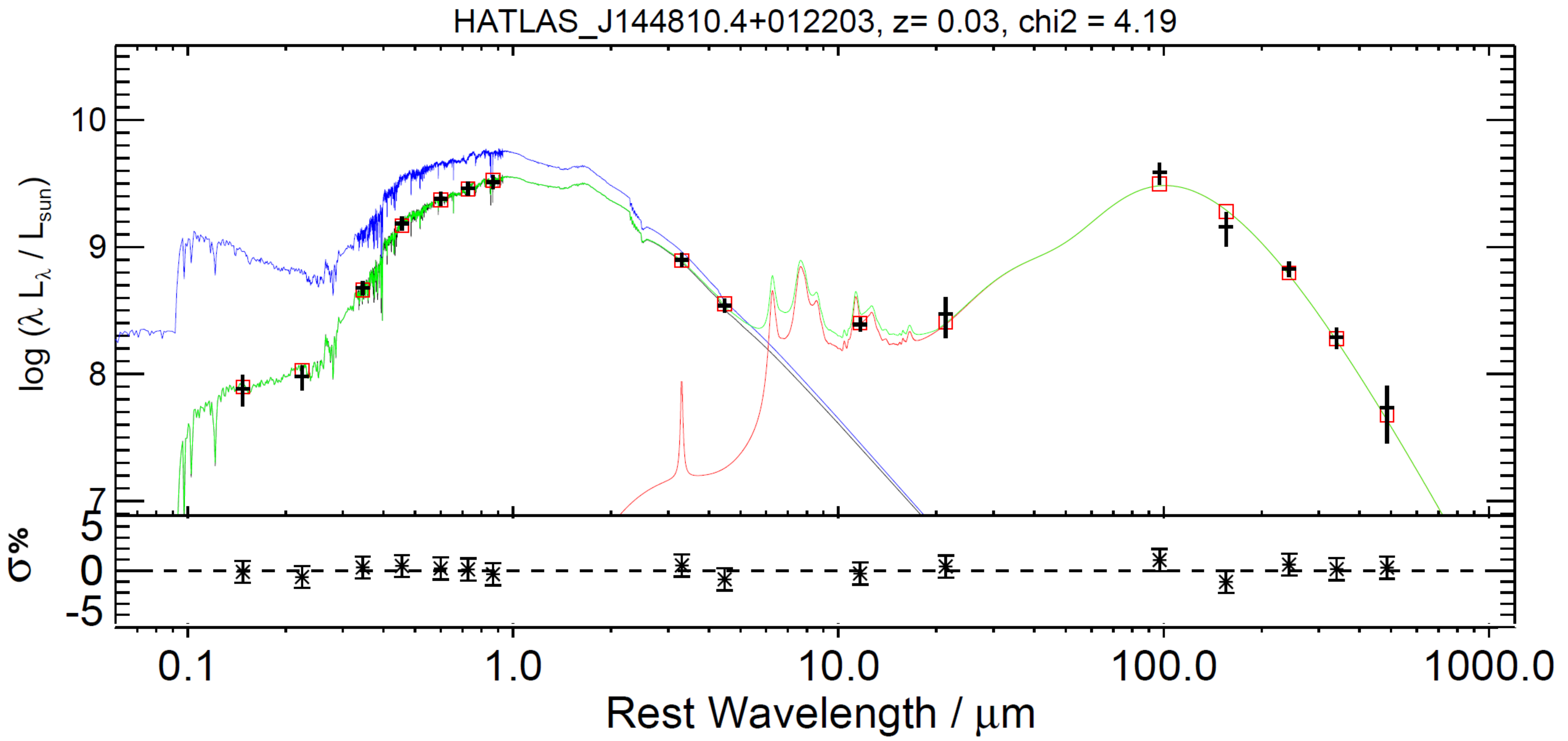}
	\caption{Top panel: A typical MAGPHYS rest-frame SED fit of an H-ATLAS \rg source.  Observed 
	UV to \submm fluxes are shown with plus symbols. The green line is the best-fitting model 
	while the blue line is the unattenuated stellar fitted spectrum.
    Bottom panel: The fit residuals $\sigma$ in per cent estimated according to 
    $(L_{\lambda}^{\rm obs} - L_{\lambda}^{\rm model}) / L_{\lambda}^{\rm obs}$, 
    where $L_{\lambda}^{\rm obs}$ and $L_{\lambda}^{\rm model}$ are the observed and 
    model fluxes in a given photometric band.}
\label{fig:sample-SED}
\end{figure*}

For the purpose of our study, we have focused on a number of galactic parameters that are inferred by fitting the observed SEDs with MAGPHYS. These are: the galactic stellar mass ($M_{*}$), the dust mass ($M_{\rm D}$), the star formation rate (SFR), and the fraction of total dust luminosity contributed by the diffuse interstellar medium (\fmu~; 0~$\leq$ \fmu ~$\leq$~1.0). Large values of \fmu~ indicate that dust is heated by the old stellar populations while lower values suggest that  ongoing star formation has a more prominent role in heating the dust. An example of a SED fit for a \submm source in our \rg sample is shown in Fig.~\ref{fig:sample-SED}. We find that the distribution of \ki2 ~ in our sources, does not show any correlation with galaxy \nuvr~ colour indices. It is worth mentioning that the  comparison of the results from MAGPHYS, with and without the MIR constraints from {\it WISE}, shows that including the {\it WISE} data modifies the output results from MAGPHYS. The inclusion of {\it WISE} data improves the fits of the SEDs and provides better estimates of some of the parameter, and notably of the SFR. For this reason, we include in the following sections only those galaxies for which {\it WISE} data are available (e.g. $\approx  2/3$ of the main sample). This in turn, reduces the size of our sample from 3782 to 2622 sources with 78 having NUV$-r\geq 4.5$ mag and therefore are \rg.

\begin{figure}
\hspace{-2.0em}
\includegraphics[scale=0.5]{\path 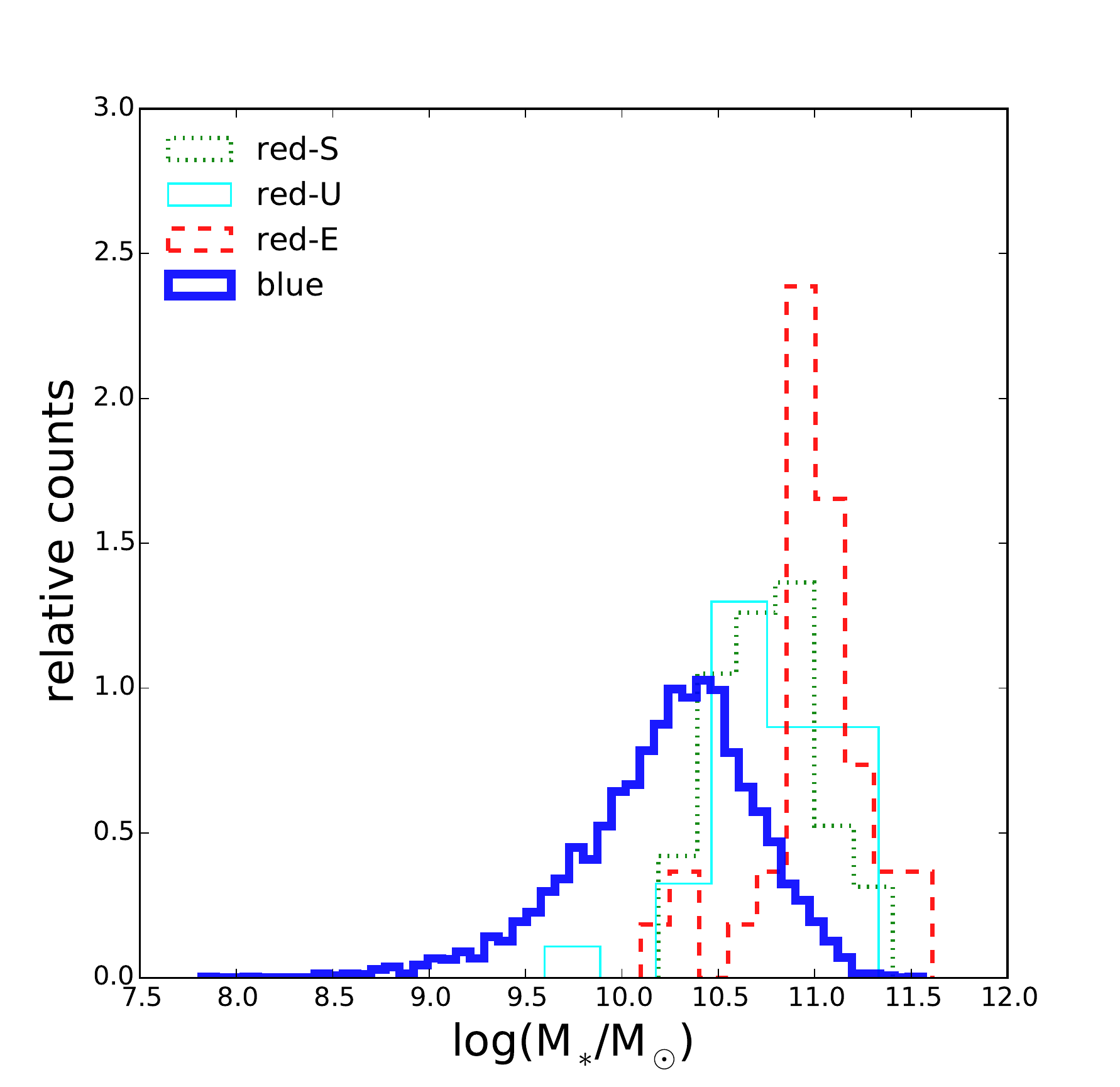}
\caption{Distribution of galaxy stellar masses in \bg(thin solid line) 
and \rg samples (\rg-S : dashed line, \rg-U : dotted line, \rg-E : thick solid line). Each histogram is normalized by its integral.}
\label{fig:fraction_of_red_gals}
\end{figure}

Figure \ref{fig:fraction_of_red_gals} displays the mass distribution of galaxies in the \bg and \rg samples (in different categories).   In our sample, $\approx$~73 per cent of \bg sources have stellar masses ${\rm log} (M_{*}/{\rm M_{\odot}}) \geq 10.0$, while the same number for the \rg galaxies is $\approx 97$~per cent, accounting for $\approx$~4.2 per cent of the total number of sources with ${\rm log}(M_{*}/{\rm M_{\odot}}) \geq 10.0$. As expected, bins associated to largest stellar masses are accupied by the \rg-E galaxies (see Table.~\ref{table:MAGPHYS-vs-HRS}).


\subsection{Dust properties}
\label{sec:MAGPHYS-dust}


It is important to compare the inferred parameters derived from MagPhys to other determinations.  We compare the estimated dust-to-stellar mass ratio ($M_{\rm D}/M_{*}$) for all sources as computed by MAGPHYS to those derived for a sample of $\sim$300 nearby galaxies from the HRS \citep[{\it Herschel} Reference Survey;][]{luca12}. The total dust mass of a given galaxy as estimated by MAGPHYS is the sum of the three components which includes the mass contributed by dust in thermal equilibrium in stellar birth clouds, as well as warm and cold dust components in the ambient interstellar medium \citep{ed08}. 

\begin{figure}
\includegraphics[scale=0.58]{\path 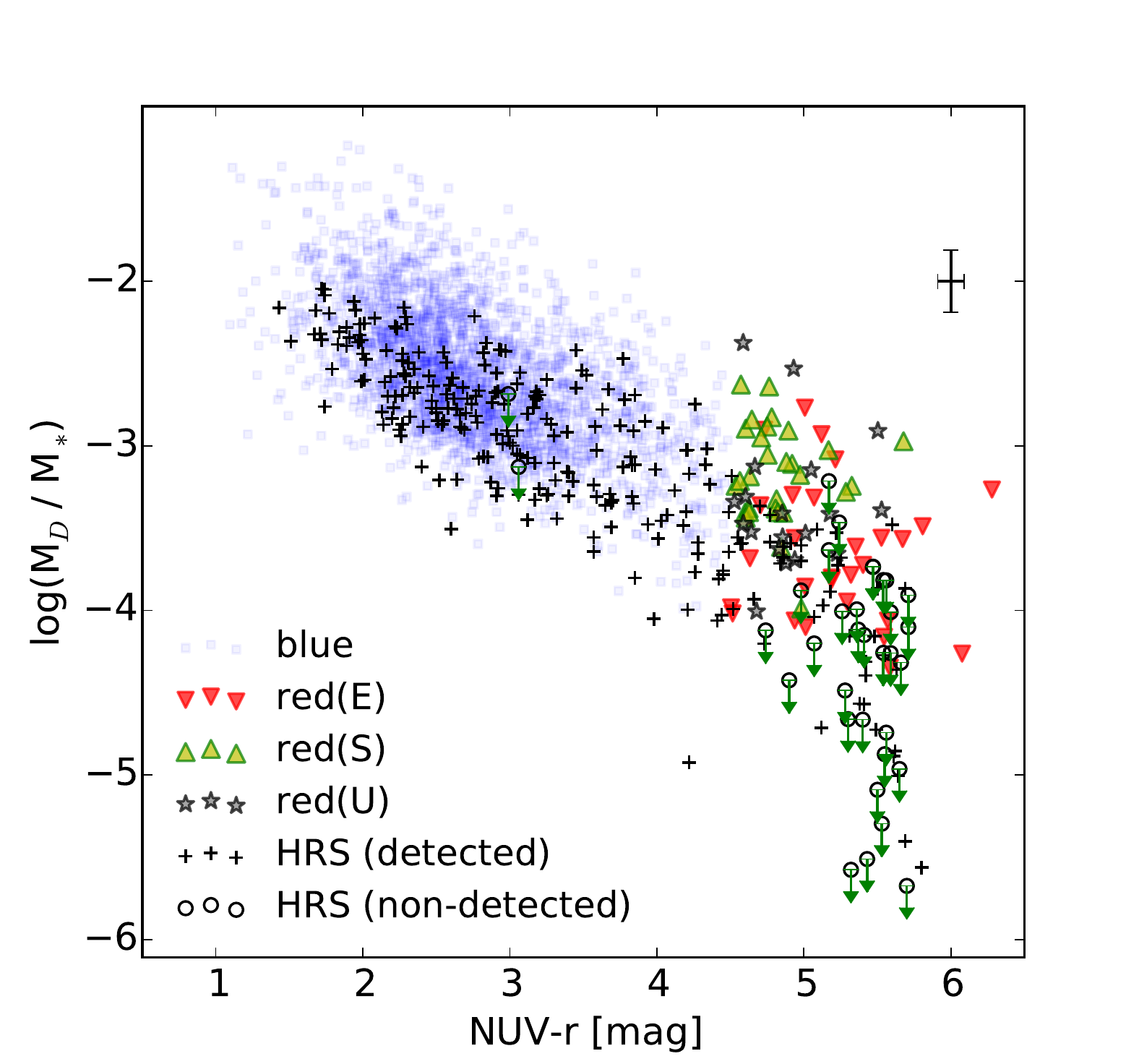}
\caption{The dust-to-stellar mass ratio as function of \nuvr~ colour
for the \bg (square) and \rg samples. E (triangle down), S (triangle up) and U (stars) 
labels represent the morphology of individual \rg source.
 The typical errors associated with our galaxies
	are indicated on the top-right corner. Overlaid are HRS 
\citep[{\it Herschel} Reference Survey;][]{luca12}	
	detected (plus sign)
	and non-detected (open circle; downward arrows indicating upper limits) galaxies.}
\label{fig:MAGPHYS-vs-HRS}
\end{figure}

Fig.~\ref{fig:MAGPHYS-vs-HRS} displays the distribution of $M_{\rm D}/M_{*}$ inferred from MAGPHYS for our sample against \nuvr~for all \rg and \bg sources. Overlaid are the $M_{\rm D}/M_{*}$ estimates from the HRS using all SPIRE bands. For HRS non-detections (triangles), the \submm upper-limit fluxes have been determined assuming a 3$\sigma$ signal over a circular aperture of radius 0.3$\times$, 0.8$\times$ and 1.4$\times$ of the optical radius for the HRS E, S0 and spirals, respectively. 

Note that in determining dust masses $M_{\rm D}$, both MAGPHYS and \citet{luca12} adopt a dust emissivity index $\beta=$~2.0 for cold dust but different dust mass absorption coefficients $\kappa_{\nu}$. \citet{luca12}  use a dust mass absorption coefficient $\kappa_{350}$ of 0.192 m$^2$kg$^{-1}$ at 350 \micron~ whereas \citet{ed08} assume  $\kappa_{850}=$~0.077 m$^2$kg$^{-1}$ at 850 \micron . Given the scaling relations $M_{\rm D} \propto \kappa_{\nu}^{-1}$ and $\kappa_{\nu} \propto \nu^{-\beta}$ one finds that $\kappa_{850}$ in MAGPHYS can be scaled down (assuming $\beta=$~2.0) to 0.45 m$^2$kg$^{-1}$ at 350 \micron~ and that 
dust masses as measured by \citet{luca12} are $\approx$~2.36 times larger than those estimated by MAGPHYS. Thus in Fig.~\ref{fig:MAGPHYS-vs-HRS}, the HRS sample are scaled down for $\approx$~0.37 dex to account for the differences between the two measurements of dust masses.

\begin{figure}
\hspace{-5.0em}
\includegraphics[scale=0.32]{\path 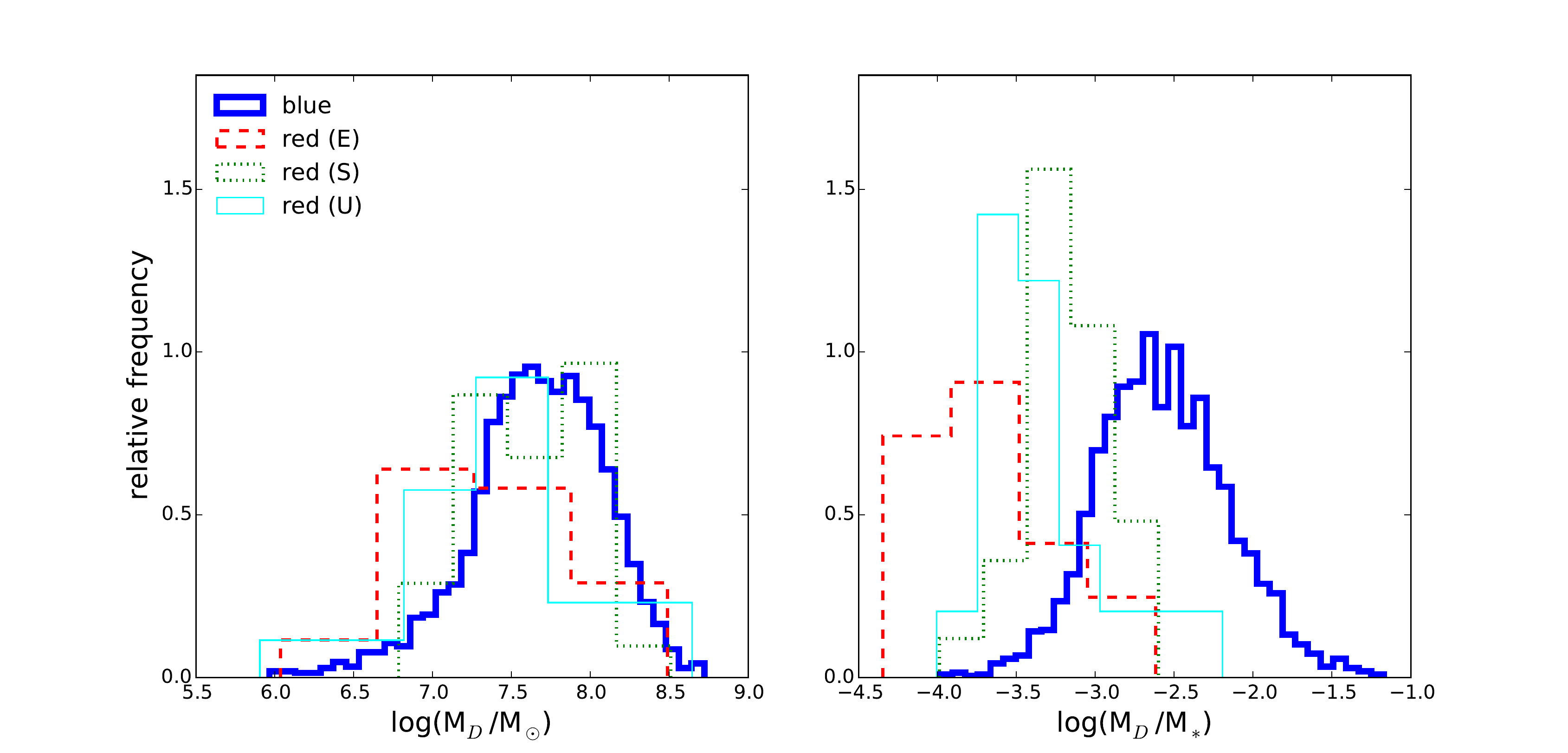}
\caption{Distributions of dust mass (left panel) as well as specific dust mass 
(right panel) in the \bg (thick solid line) and \rg sources.
E (red dashed line), S (green dotted line) and U (cyan line) 
labels represent the morphology of individual \rg source. 
Each histogram is normalized by its integral. The estimated mean value
associated to each histogram is given in Table.\ref{table:MAGPHYS-vs-HRS}.}
\label{fig:histo-Mdust}
\end{figure}

It can be seen that the $M_{\rm D}/M_{*}$ ratios for both the \bg or \rg galaxies agrees reasonably well with estimates from the HRS detected objects. Furthermore, the \rg sources of type E exhibit, on average, $M_{\rm D}/M_{*}$ ratios  that are noticeably lower than those of \bg galaxies. This is even more clear in the right panel of Fig.\ref{fig:histo-Mdust} which displays the distributions $M_{\rm D}/M_{*}$ in all sources. The mean values as summarized in Table.~\ref{table:MAGPHYS-vs-HRS} suggest that the \rg-E objects have values of the dust-to-stellar masses that are approximately an order of magnitude lower than those in the \bg sources. This is partly because the \rg-Es have high stellar masses but as is visible in the left panel of Fig.~\ref{fig:histo-Mdust}, they also have a lower dust content in comparison to the \rg-S and \bg systems.
Note that the distribution of specific dust mass of the \rg-⁠S galaxies
does not match the distribution of the \bg star forming galaxies. We
will discuss this further in Sec.~\ref{sec:MAGPHYS-sfr}.
\begin{center}
\begin{table*}
\begin{tabular}{lcccccc}

Galaxy type & $\log$(SFR)[M$_{\odot}$yr $^{-1}$] & $\log$(SFR/$\rm{M_*}$)[yr $^{-1}$] & $\log$(\mdust) & $\log$(\mstar) & $\log$(\Mdust) & f$_{\mu}$ \\ \hline

\bg 			&  0.43$\pm$0.57		& $-9.72\pm$0.80		& 7.84$\pm$0.54		& 	10.42$\pm$0.47		& 	$-2.58\pm$0.62		& 	0.55$\pm$0.53	\\
\rg (type-S)	& $-0.29\pm$0.54		&$-11.11\pm$	0.65		& 7.74$\pm$0.44		& 	10.86$\pm$0.37		& 	$-3.12\pm$0.51		&	0.88	$\pm$0.31	\\
\rg (type-U)	& $-0.71\pm$0.53		&$-11.34\pm$0.53		& 7.67$\pm$0.39		& 	10.83$\pm$0.29		&   $-3.16\pm$0.44		&	0.88	$\pm$0.22	\\
\rg (type-E)	& $-0.67\pm$0.63		&$-11.70\pm$0.62		& 7.62$\pm$0.49		&   11.06$\pm$0.26		&   $-3.44\pm$0.51		&   0.92$\pm$0.29	\\ \hline

\end{tabular}
\caption{Mean values of various MAGPHYS output parameters estimated from distributions 
shown in  Figs.\ref{fig:histo-Mdust} and \ref{fig:MAGPHYS-sfr-3-histograms}.}
  \label{table:MAGPHYS-vs-HRS}
  \end{table*}
\end{center}


\begin{center}
\begin{table*}
\begin{tabular}{lcccccc}

Parameter  & \bg vs. \rg-E &  \bg vs. \rg-S &  \bg vs. \rg-U & \rg-E vs. \rg-S &  \rg-E vs. \rg-U &   \rg-S vs .\rg-U      \\ \hline

S\'{e}rsic  index                      & $\bf<0.001$ & $\bf<0.001$ & $\bf<0.001$ & $\bf<0.001$ & 0.098         & 0.056        \\
 ellipticity                           & $\bf<0.001$ & 0.40          & 0.0045        & $\bf<0.001$ & 0.013         & $\bf<0.001$\\
 $\log(\Sigma_{5})$                    & 0.25          & 0.43          & 0.13          & 0.71          & 0.46          & 0.63         \\
 $\log$(\mstar)                        & $\bf<0.001$ & $\bf<0.001$ & $\bf<0.001$ & $\bf<0.001$ & $\bf<0.001$ & 0.94         \\
 $\log$(SFR)[M$_{\odot}$yr $^{-1}$]    & $\bf<0.001$ & $\bf<0.001$ & $\bf<0.001$ & 0.021         & 0.87          & 0.032        \\
 $\log$(SFR/$\rm{M_*}$)[yr $^{-1}$]    & $\bf<0.001$ & $\bf<0.001$ & $\bf<0.001$ & $\bf<0.001$ & 0.10          & 0.012        \\
 $\log$(\mdust)                        & 0.50          & 0.0049        & 0.0021        & 0.029         & 0.89          & 0.23         \\
 $\log$(\Mdust)                        & $\bf<0.001$ & $\bf<0.001$ & $\bf<0.001$ & $\bf<0.001$ & 0.014         & 0.012        \\
 f$_{\mu}$                             & $\bf<0.001$ & $\bf<0.001$ & $\bf<0.001$ & 0.056         & 0.87          & 0.45         \\ \hline
\end{tabular}
\caption{The results of a Kolmogorov-Smirnov test ($p$-values) associated to parameter distributions  
shown in  Figs.\ref{fig:red-morphology}, \ref{fig:density-sigma5}, \ref{fig:fraction_of_red_gals},  \ref{fig:histo-Mdust} and \ref{fig:MAGPHYS-sfr-3-histograms}. {\revised We highlight with bold face fonts those parameters for which the KS-test indicates a significant difference in the underlying distributions, i.e. $p<0.001$.}}
  \label{table:kstest}
  \end{table*}
\end{center}

\subsection{Star formation rates}
\label{sec:MAGPHYS-sfr}

\begin{figure}
\hspace{-2.0em}
\includegraphics[scale=0.5]{\path 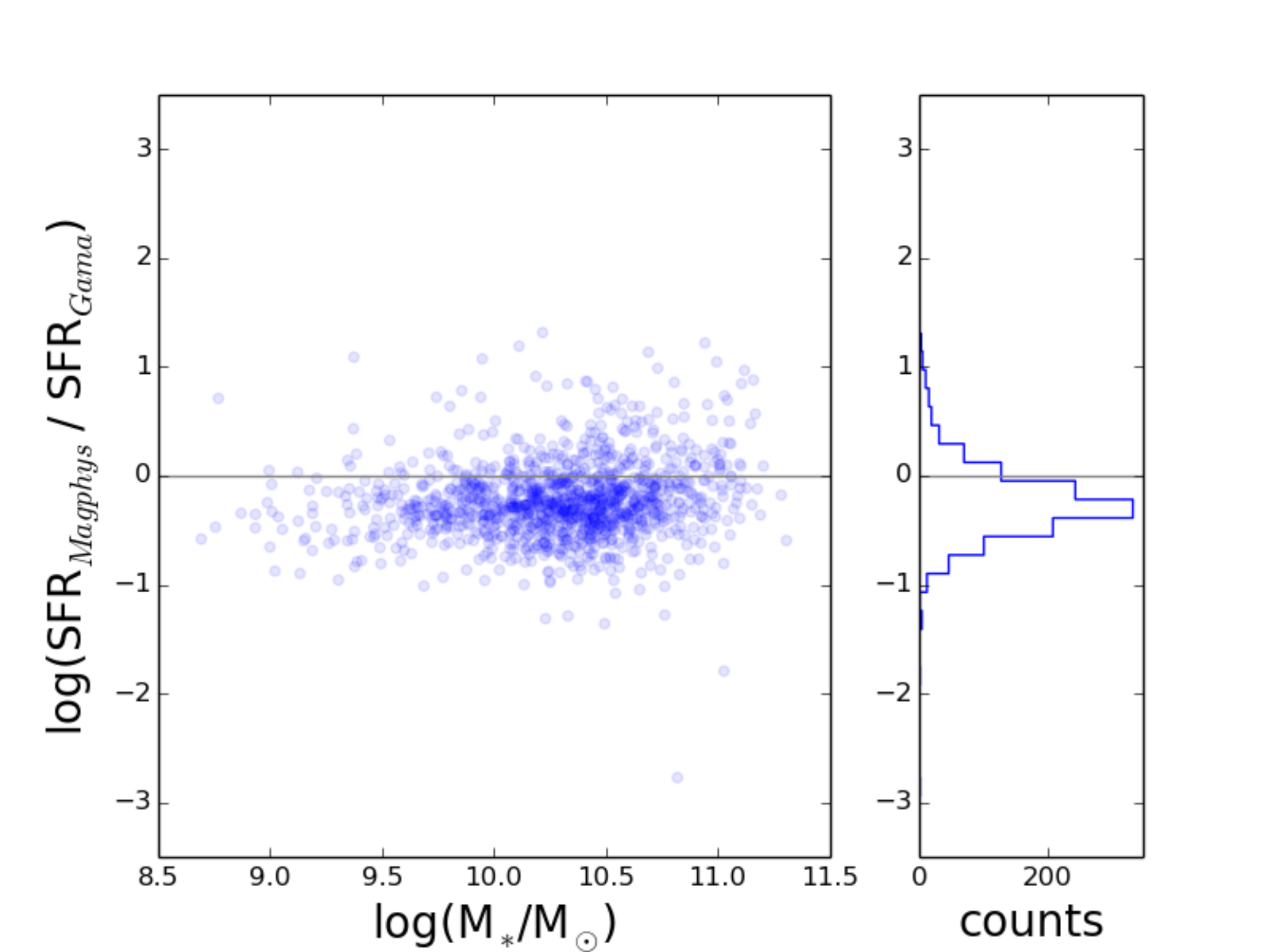}
\caption{ Ratio of MAGPHYS SFR over GAMA DR2 SFR in logarithmic scale vs \M* for all galaxies in our sample (see Eq.~\ref{eq:sfr}). Vertical histogram shows the distributions of data points along y-axis.}
\label{fig:MAGPHYS-sfr}
\end{figure}

In Fig.~\ref{fig:MAGPHYS-sfr}, we compare  the MAGPHYS derived values of the star formation rates (SFRs) to those estimated based on the spectral analysis of the H$\rm \alpha$ lines using the Second GAMA Data Release (GAMA-DR2) catalogues \citep{whb,hopkins13,gunawardhana13,liske14}.


Galaxy SFRs in GAMA-DR2 are determined from the \citet{kennicutt98} relation and based on the  total aperture-corrected H$\rm \alpha$ luminosities observed through fibre spectroscopy. The $r$-band absolute magnitude of each galaxy have been used in order to correct for the Aperture and therefore recovering the total H$\rm \alpha$ luminosities \citep{hopkins03,gunawardhana11}. Dust corrections were estimated for each galaxy from  the observed Balmer decrement. Finally stellar absorption corrections were applied to both H$\rm \alpha$ and H$\rm \beta$ fluxes which together with the H$\rm \alpha$ equivalent width (EW) allow to calculate the total aperture-corrected H$\rm \alpha$ luminosities as described in detail in \citet{hopkins03}.

We find a strong correlation between the two estimates of SFRs such that (SFRs are in units of $\rm{M_{\odot} yr^{-1}}$)

\begin{equation}
\log {\rm SFR_{Magphys}} = 1.22\substack{+0.02 \\ -0.02}\times \log {\rm SFR_{GamaDR2}} - 0.35.
\label{eq:sfr}
\end{equation}
Give the Pearson correlation coefficient of $r\simeq$~0.71 in the above equation, it is evident that in general,  GAMA DR2 H$\rm \alpha$ -derived SFRs are well correlated  with those predicted by MAGPHYS through SED based measurements though on average MAGPHYS derived SFRs are $\approx$~0.3 dex lower than those based on the H$\rm \alpha$ luminosities from GAMA. This may be due to different treatments applied in correcting for dust or aperture as explained in \citet{wijesinghe11}.

The distribution of SFR related parameters are displayed in Fig.~\ref{fig:MAGPHYS-sfr-3-histograms}. The first two panels, show the SFR and the specific star formation rate (SSFR) of \bg and \rg galaxies. The mean value of the SFR in the \rg-E galaxies is an order of magnitude lower than in the \bg galaxies with SFR$_{\bg}$/SFR$_{\rg{\rm-E}} \approx$13 (SFR$_{\rg{\rm-S}}$/SFR$_{\rg{\rm-E}} \approx$2.5 ; see also Table.~\ref{table:MAGPHYS-vs-HRS}).

The difference between the two samples is even more pronounced when considering SFR normalized by galaxy's stellar mass $M_{*}$ such that SSFR$_{\bg}$/SSFR$_{\rg{\rm-E}} \approx$100 (SSFR$_{\rg{\rm-S}}$/SSFR$_{\rg{\rm-E}} \approx$4).  For both the SFR and the SSFR, the values estimated for the \rg-S type sources and the galaxies with uncertain morphology, lay between the \rg-E galaxies and the \bg control sample. In comparison, \citet{kr12} (i.e. Table C1) measure $-9.99\substack{+0.03 \\ -0.03}$ and $-10.85\substack{+0.14 \\ -0.14}$ for SSFR in samples of 'H-ATLAS spiral' and 'H-ATLAS elliptical' galaxies respectively.

Figure.~\ref{fig:MAGPHYS-sfr-3-histograms}c shows the normalized distributions of \fmu~ in the \bg and \rg populations. The \rg-E galaxies have an average \fmu$\sim$0.92, well above the mean ($\sim$0.55) of the \bg galaxies. This indicates that while about half of the observed FIR emission observed in the \bg galaxies comes from dust in birth clouds, the FIR of \rg-E galaxies is dominated by dust in the diffuse interstellar medium (ISM). We note that the average derived \fmu~ for the \rg-S systems is significantly higher than for the \bg control sample and only slightly lower than for the sample of the \rg-E galaxies.

\subsubsection{On the derived properties of the \rg-⁠S sample}
\label{subsub:sfr-red-s}

\begin{figure*}
\hspace{-10em}
\includegraphics[scale=0.31]{\path 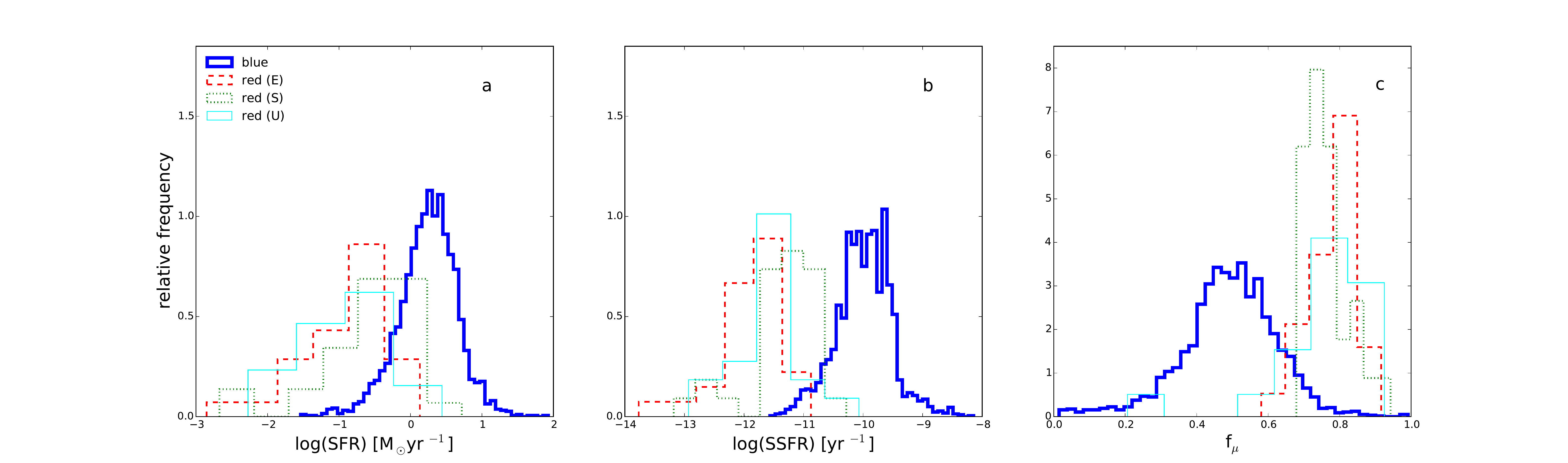}
  \caption{Distributions of (a) star formation rate, (b) specific star
    formation rate and (c) \fmu, e.g. the fraction of total dust
    luminosity contributed by the diffuse interstellar medium, in the \bg
    (thick solid line) and the \rg sources. E (red dashed line), S (green
    dotted line) and U (cyan line) labels represent the morphology of
    individual \rg source. Each histogram is normalized by its
    integral. The estimated mean value associated to each histogram is
    given in Table.\ref{table:MAGPHYS-vs-HRS}.}
\label{fig:MAGPHYS-sfr-3-histograms}
\end{figure*}

Even though the \rg-⁠S galaxies are not the prime focus of this paper, this sample does display some interesting characteristics that are worth commenting on briefly. As can be derived from figures \ref{fig:fraction_of_red_gals}, \ref{fig:histo-Mdust}b and \ref{fig:MAGPHYS-sfr-3-histograms} the deduced properties of the \rg-⁠S galaxies do not match the \bg galaxy properties. The \rg-⁠S galaxies appear intermediate between the \rg-⁠E and the \bg galaxies in stellar mass, SFR and specific dust mass. This offset is primarily driven by the higher derived stellar masses and the correspondingly lower SFR. This is contrary to what one would expected if the red colours of the edge-⁠on galaxies are \emph{only} due to their high inclination.

\begin{figure}
\includegraphics[scale=0.55]{\path 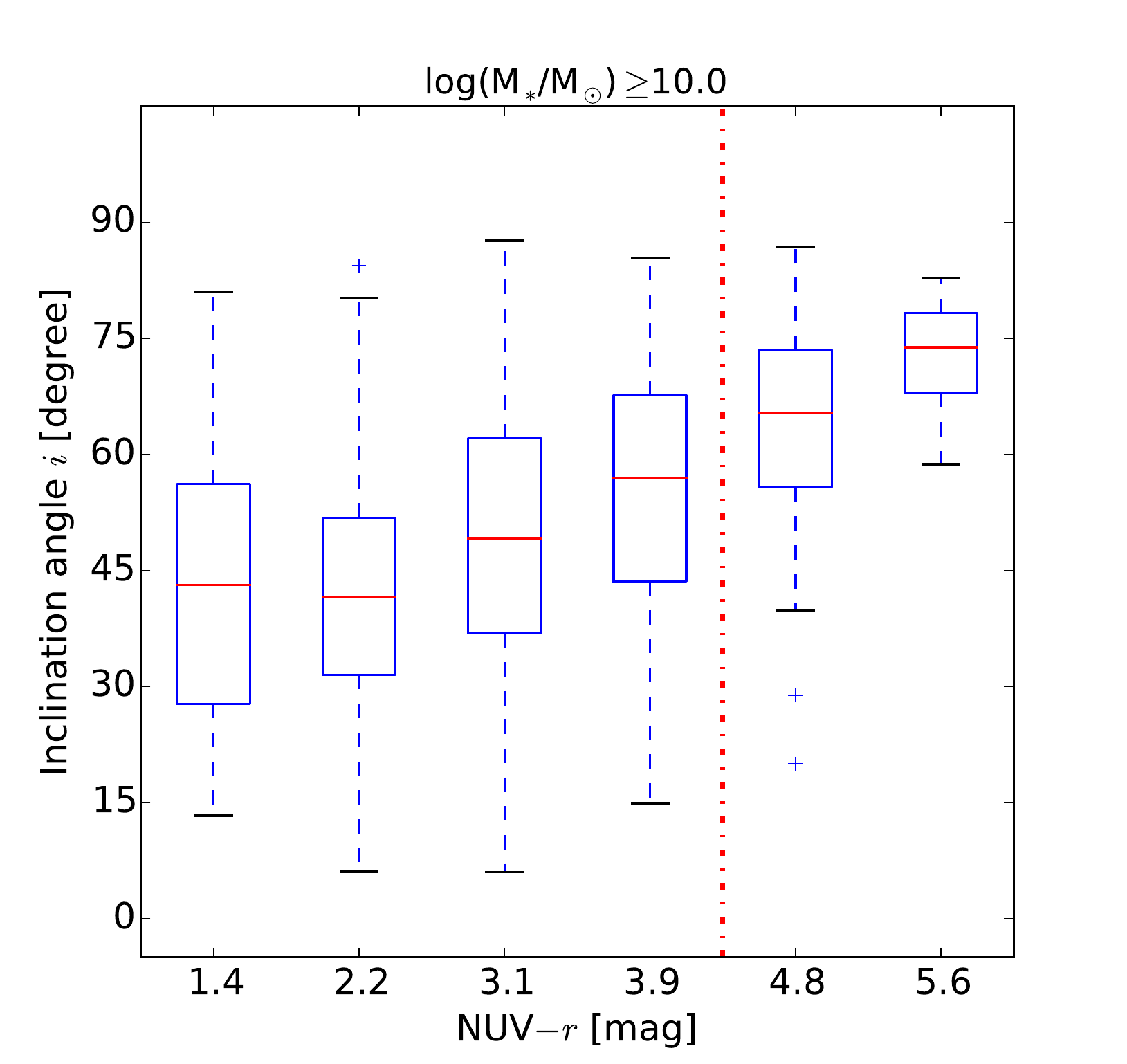}
\caption{ Distribution of galactic inclination angles $i$ for \bg and  \rg-S galaxies, having stellar masses ${\rm log}(M_{*}/{\rm M_{\odot}}) \geq 10.0$, vs. NUV$- r$ colour. Each box extends from the lower to upper quartile values of data, with a line at the median (red line). Inclination angles are computed using Eq.~\ref{Eq: incl_angle}.  Dashed lines extending vertically from the boxes indicating variability outside the upper and lower quartiles. Individual data points indicate outliers. The vertical dashed-dotted line intersects the x-axis at NUV$-r = $~4.5 above which galaxies are classified as \rg. }
\label{fig:boxplot}
\end{figure}

Inclination does play a significant role in defining this sample, as can be concluded from Fig.~\ref{fig:boxplot}. We show in this figure the inclination of the \bg $+$ \rg-⁠S~ for the stellar masses above ${\rm log}(M_{*}/{\rm M_{\odot}}) \approx 10.0$, i.e. the range of stellar masses of interest. There is a definite trend of the median inclination against observed optical redness and in particular the very reddest sources are almost exclusively very inclined sources.

We see two main interpretations ---⁠ which could be at play simultaneously --- that could explain these characteristics of the \rg-⁠S~ sample.

\begin{enumerate}[(i)]
\item High inclination is a necessary, but not sufficient condition for a star-forming disk galaxy to be \submm detected {\emph {and}} very optically red. In this case the red colour would apparently select preferentially the more massive disk galaxies. Perhaps the less massive disk galaxies have enough star formation in their periphery of their disks ---⁠ which would not be strongly obscured, even in the case of strong inclination --- to exhibit a blue-⁠ish optical colour. Alternatively, the red colour of those massive disks could be a direct results of a dominant old stellar population. \\

\item The galaxy parameters, derived from MAGPHYS, of the very inclined and dusty sources are systematically biased to higher stellar masses and less star formation. This is in line with the finding of \citet{dc10}. These authors find that the derived SFR for edge-⁠on galaxies is about a factor 3$\times$ ($\approx 0.48$~dex) below their face-⁠on counterparts.  They also find that this effect is also responsible for the lower dust masses (or dust luminosities) and higher \fmu~ estimated for edge-on in comparison to face-on galaxies. The amplitude of this effect is insufficient to directly explain the difference we find between the blue sample and the \rg-⁠S sample. Note however that \citet{dc10} describe the effect on an {\it inclined} sample of galaxies while the \rg-⁠S sample is selected to have only galaxies with very red colours.  The {\it inclined} sample  contains galaxies with varying degrees of hidden star-formation whereas the \rg-⁠S sample contains only galaxies with very obscured star-formation. We thus would expect to find a larger offset of the derived parameters in the \rg-⁠S sample than in the {\it inclined} sample.

\end{enumerate}
Clearly this red disk population of nearby galaxies deserves further attention in a dedicated study.

\subsection{Dust mass correlations with galactic properties}
\label{sec:source-of-dust}

\begin{figure}
\vspace*{-5mm}
\includegraphics[scale=0.35]{\path 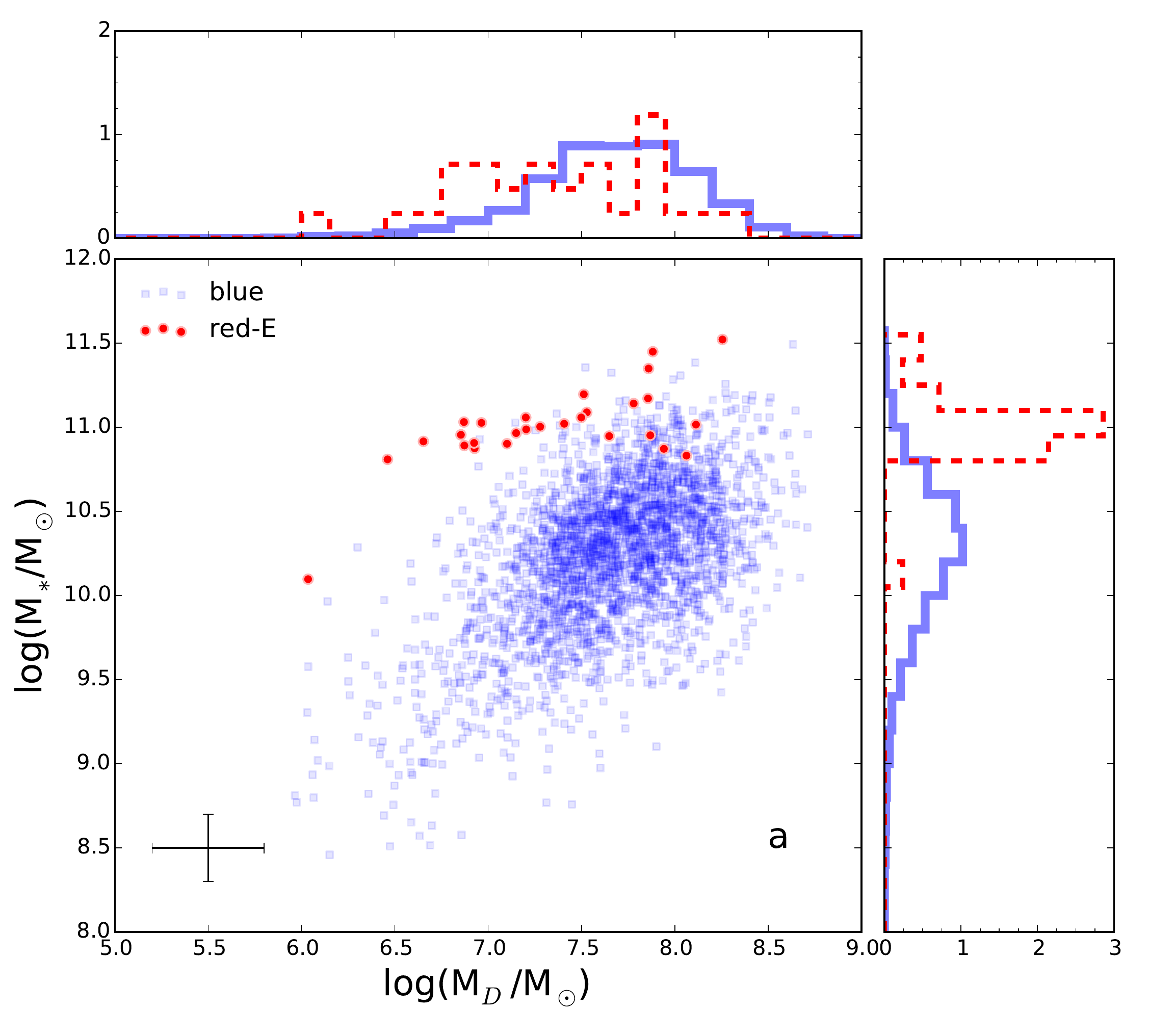}
\includegraphics[scale=0.35]{\path 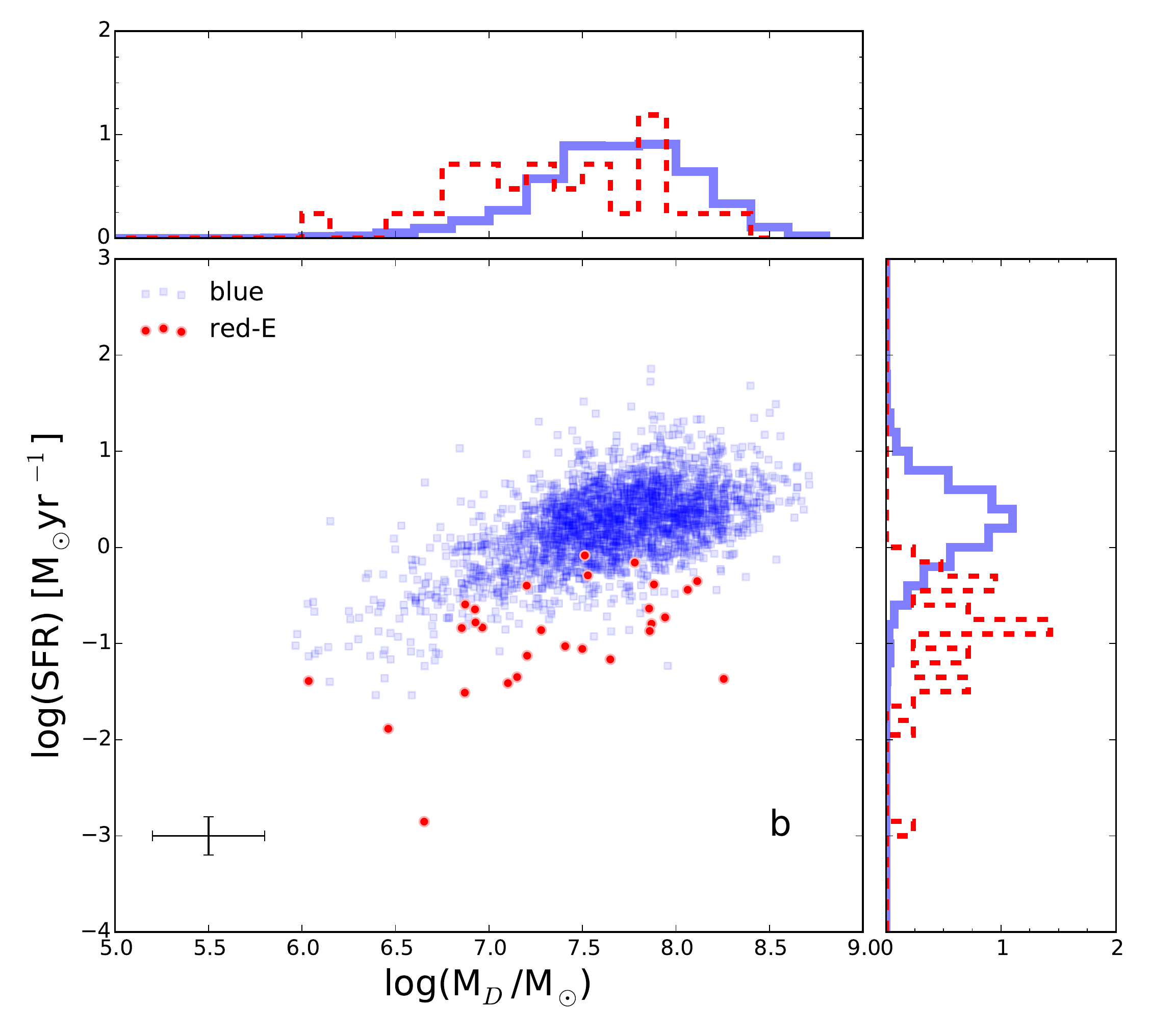}
\includegraphics[scale=0.35]{\path 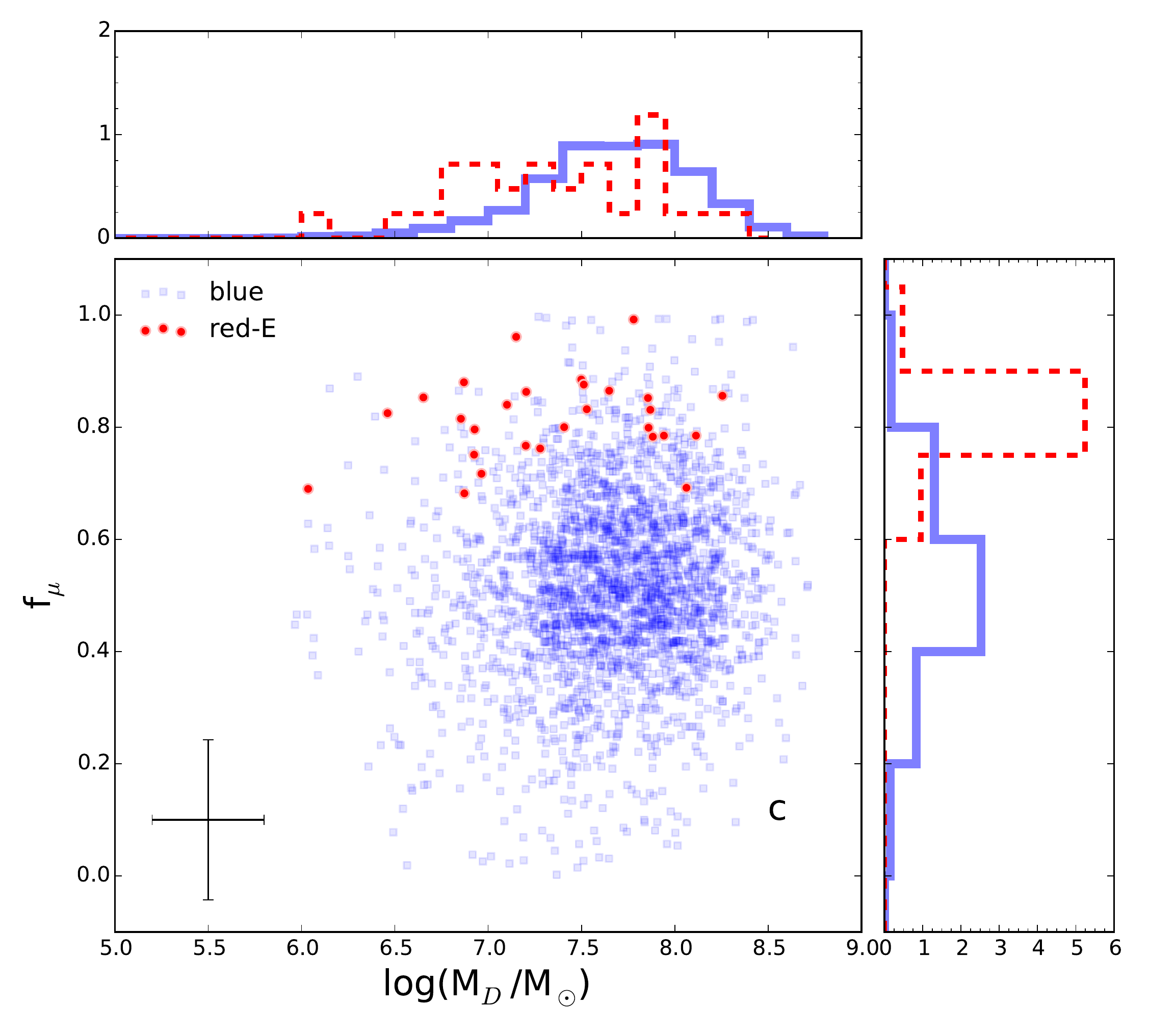}

  \caption{Distribution of dust mass \mmdust~ against (a) stellar mass \M*
  (b) star formation rate SFR and (c) \fmu~ in \bg (blue square), 
    \rg-E (red circle). In addition, horizontal and vertical
    histograms show the distributions of data points along x and y
    axes with blue/thick and red/dashed lines representing \bg and 
    \rg-E. Each histogram is normalized by its integral.
    Typical errors associated to various parameters are indicated on
    the bottom-left corner. Results of linear regression analysis to \bg and \rg-E observed data points in panels 'a' and 'b'  are given in Table.~\ref{table: linaer-fit-parameters}.}
\label{fig:3panels}
\end{figure}

We show in Fig.~\ref{fig:3panels} correlation plots of the derived dust mass versus a number of key parameters ($M_{*}$ , SFR, and \fmu) in the \rg-E and \bg galaxies. These parameters have been chosen to elucidate the possible origin and role of the dust in the \rg-E galaxies. The first conclusion that can be drawn from the perusal of these diagrams is that the \rg-E galaxies clearly occupy a different parameter space from \bg spiral galaxies. 

Panel~\ref{fig:3panels}a shows a very different behaviour of the $M_{\rm D}$ as a function of $M_{*}$ for the blue galaxies and the \rg-⁠E sample. The blue sample shows a roughly linear correlation (with scatter) between the dust reservoir and the $M_{*}$. This relation is expected due to the $M_{*}$-⁠SFR relation for normal galaxies, if the $M_{\rm D}$ is measuring the reservoir available for star formation. The \rg-⁠E sample exhibits a totally different behaviour apart from being located in a distinctly different part of this diagram. While the host galaxies are all -⁠-⁠ with one outlier -⁠-⁠ of very similar mass ($\approx$~10$^{11}$M$_{\odot}$) their dust content spans more than two orders of magnitude. This complete decorrelation of stellar mass and dust content argues against a stellar origin \citep[e.g. ][]{luca12} for the dust in those galaxies. While for blue galaxies the dust mass increases with stellar mass, the dust masses found for the \rg-E span $\approx 2$ order of magnitudes for stellar masses that are roughly constant at $\approx$ $10^{11}$ M$_{\odot}$ (see Table.~\ref{table: linaer-fit-parameters}). 

In panel~\ref{fig:3panels}b we show that there {\emph is } a moderate correlation in the \rg-E galaxies between the derived SFR and $M_{\rm D}$ with a similar slope but offset from the \bg sequence. We interpret the existence of this correlation as an indication that the star formation is probably taking place in the cold gas associated with the dust. 

The observed offset between the \bg control sample and the \rg-E sample implies that the same amount of dust in the \rg-E galaxies is associated with about an order of magnitude less star formation. This could be an indication that the physical state of the cold ISM phase in the \rg-E galaxies is significantly different perhaps due to the very different environment in which the cold gas is embedded. This interpretation is corroborated by panel~\ref{fig:3panels}c where we show that indeed the MAGPHYS derived fraction of the dust heating due to the interstellar radiation field, i.e. \fmu~ is much higher in the \rg-E galaxies than their \bg counterparts.

\begin{table*}
\begin{tabular}{lllccccr}\hline

Galaxy            & Y  & X   & $s$       & $\pm$err           & $c$           & $r$-value             & $p$-value \\
type              &    &     & (slope)   & standard deviation & (intercept)   & (Pearson correlation) &           \\ \hline\hline

\bg   (panel $a$) & $\log({\rm M_*/M_{\odot}})$          & $\log({\rm M_D/M_{\odot}})$  & 0.56 & 0.01 &  5.93 & 0.54 & $<$0.001 \\
\rg-E (panel $a$) & -                                    & -                            & 0.30 & 0.06 &  8.76 & 0.67 & $<$0.001 \\ \hline
\bg   (panel $b$) & $\log$(SFR) [M$_{\odot}$yr $^{-1}$]  & $\log({\rm M_D/M_{\odot}})$  & 0.54 & 0.01 & -3.89 & 0.56 & $<$0.001 \\
\rg-E (panel $b$) & -                                    & -                            & 0.54 & 0.18 & -4.87 & 0.50 & 0.006    \\ \hline

\end{tabular}
\caption {Results of linear regression analysis to the observed data points in panels 'a' and 'b' of Figs.~\ref{fig:3panels}. Parameters in the table are associated to the linear model ${\rm Y} = s(\pm {\rm err}) \times {\rm X} + c$.}

\label{table: linaer-fit-parameters}
\end{table*}

\subsection{The origin of dust in \rg-E}

In the classical definition of galactic types, ellipticals were classified as devoid of gas and dust \citep{hubble26, devaucouleurs59,sandage61}. In the subsequent years, dust emission in Ellipticals has been detected from the ground \citep{hawarden81,sadler85,sparks85,kormendy87,ebneter88, pandey01} and from space using the {\it Infrared Astronomical Satellite (IRAS)} \citep{jura87,knapp89} and the {\it Spitzer Space Telescope} \citep{roccavolmerange07}. Dust lanes were observed early on along the minor axis of ellipticals \citep{bertola78}. When in some ellipticals the dust lanes and stars were observed to rotate in opposite direction, this was suggestive that this dust must have been accreted and can not be accounted for by mass loss from evolved stars \citep{kormendy89}. Kinematic information is important in order to constrain the presence of counter-rotating gas (and dust) in ellipticals in order to establish the frequency of the accretion scenario \citep{bertola88}   
 
In this study, the unresolved red ellipticals detected in the \submm do not have associated kinematic information. However, we attempt to establish whether the present dust masses in our sample of elliptical galaxies can be explained with stellar sources using a model of dust formation and evolution in ellipticals. We compare the specific dust masses ($M_{\rm D}/M_{*}$) with the predictions for dust mass return from a single stellar population (SSP) model and which represents an instantaneous burst of star formation. {\revised The star formation histories of the observed galaxies are more complex than that represented by a single burst of star formation. Their stellar masses and colours are however clearly dominated by the old stellar populations. Moreover, chemical evolution models of elliptical galaxies find very short timescales of their formation and high star formation efficiencies of the initial starburst \citep{Pipino05}. The present SFR of $\sim 0.1 \rm{M_{\odot}yr^{-1}}$ in our sample is several orders of magnitude lower than that the SFR in the past responsible for the build-up of their stellar mass of $\sim 10^{11}\rm{M_{\odot}}$. Therefore, for comparison with the dust model predictions, we assume that the entire stellar mass of each red-E galaxy is associated with a single burst with an age equal to its mass weighted age derived from the SED fitting. The observed dust mass in a galaxy is thus compared with the survived dust mass from the SSP with the same age.} The model of the SSP adopted here was introduced in \cite{zhukovska08} and was used to describe the chemical evolution of dust and gas in the Milky Way and dwarf galaxies \citep{zhukovskaetal08, zhukovska14}. For the chemical evolution aspects of the SSP model, we adopt the same ingredients as in \cite{zhukovska08} except for the IMF, for which we use the \cite{Chabrier03} form. This is consistent with the IMF that is adopted in the SED fitting with MAGPHYS.  

\begin{figure*}
\includegraphics[width=\textwidth]{\path 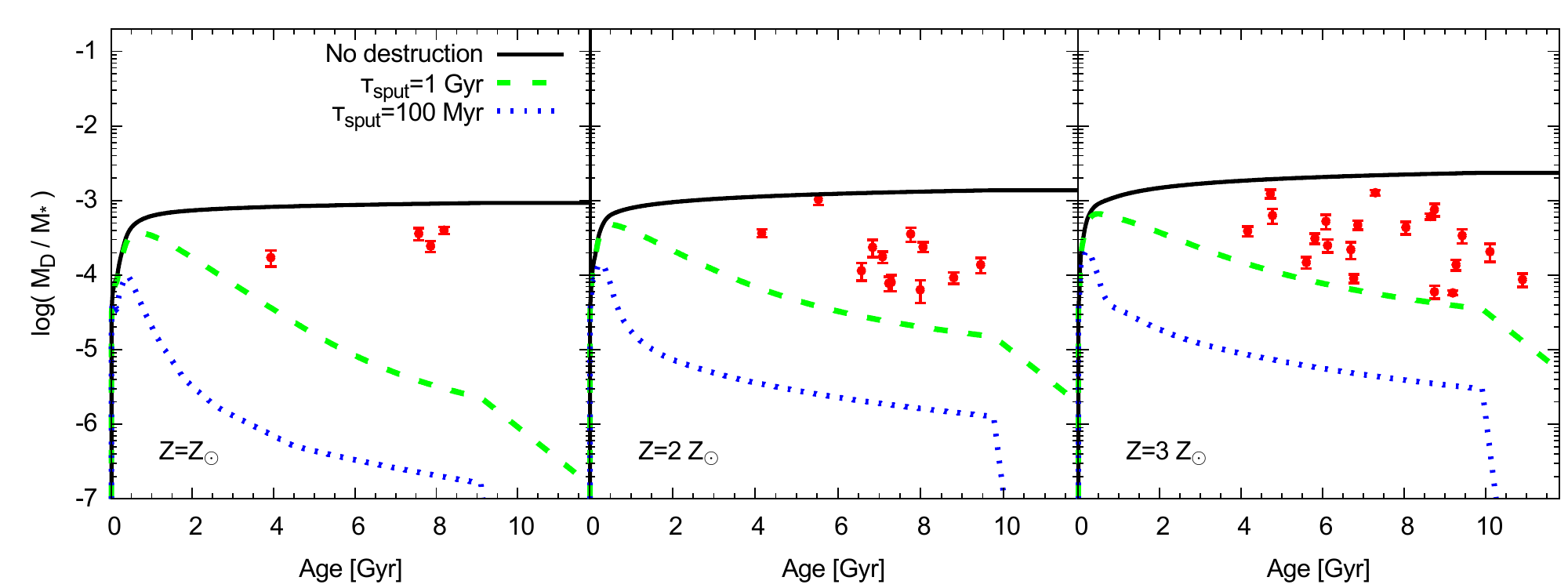}
\caption{The evolution of the dust mass relatively to the stellar mass of as a function of the age of single stellar population. The left, middle and right panel indicate initial metallicities of $Z=$ Z$_{\odot}$,  $Z=2$ Z$_{\odot}$ and  $Z=3$ Z$_{\odot}$, respectively. The value of the solar metallicity adopted here is Z$_{\odot}=0.014$ \citep{asplund09}. The solid lines shows the evolution of the cumulative dust mass returned in the SSP. The evolution of dust mass for the same SSP model with dust destruction by thermal sputtering on the timescales of 1~Gyr and 100~Myr are shown with the dashed and dotted lines, respectively. The filled red circles represent the sample of \rg-E galaxies which have been grouped in metallicity bins of [$0.5-1.5$] Z$_{\odot}$, [1.5-2.5] Z$_{\odot}$, and $> 2.5$ Z$_{\odot}$. The specific dust masses of each \rg-E galaxy in the sample is plotted versus the mass weighted age of its stellar populations and the metallicity of each galaxy is obtained from the SDSS DR4 \citep{gallazi05}.}
\label{dustmodel}
\end{figure*}

The model includes dust production by type II supernovae (SNe) and by asymptotic giant branch (AGB) stars. Type Ia SNe are an important source of metallic iron in early type galaxies. Models of dust evolution imply that, with an assumption of high condensation efficiencies of metals into dust in their ejecta, they can dominate dust input in elliptical galaxies \citep[e.g.,][]{calura08,pipino11}. Far-infrared observational surveys of both warm and cool dust in remnants of type Ia SNe do not however find evidence of efficient dust formation, in contrast to remnants of type II SNe \citep{gomez12}. This is supported by theoretical models, which indicate that newly formed grains are small and are easily destroyed in shocked gas before being ejected into the ISM \citep{noz11}. Therefore, we neglect the dust input from type Ia SNe. 

The net input from type II SNe is still debated. We add their contribution for completeness, as they produce dust for a limited period of time after stars have formed ($\sim 40$ Myr). We adopt relatively low efficiencies of dust condensation in the SNe ejecta. These are constrained by meteoritic data and the observed metallicity-dust to gas ratio relation in dwarf galaxies \citep{zhukovskaetal08,zhukovska14}. 

The mass- and metallicity-dependent dust yields for AGB stars are taken from the work of \cite{ferrarotti06} with additional models from \cite{zhukovskaetal08}. These dust yields were computed for stellar metallicity ranging from $Z=0.001$ up to the suprasolar values of $0.04$ and for the stellar mass range $[1-7]$ M$_{\odot}$. We extrapolate the dust yields in the mass range $[7-8]$M$_{\odot}$. Only one galaxy in the red-E sample is old enough for stars with masses below 1~M$_{\odot}$ to contribute to the dust budget. However, stars in this mass range loose a large fraction of their envelopes during Red Giant Branch evolution characterised by  inefficient dust formation \citep{Gail2009p512, McDonald2011hz, McDonald2015fy}. Some amount of dust is condensed during following AGB stage, but the total dust mass returned to the ISM is very low. Estimates based on the gas mass-loss rates derived in \cite{McDonald2011eo} and \cite{McDonald2015fy} point to $\lesssim 10^{-3}$M$_{\odot}$ of dust per star. Given this low value, we choose not to extrapolate the dust yields down to 0.8~M$_{\odot}$ and neglect dust input from these stars.

The ISM in elliptical galaxies is dominated by hot rarefied gas with temperatures of $\sim 10^7$~K \citep{mathews03}. Grains can be rapidly sputtered in high-temperature gas due to collisions with ions (mostly with abundant H$^+$) \citep{draine79,itoh89}. The time scale of destruction by thermal sputtering can be approximated as

\begin{equation}
\tau_{\rm sput}= 10^5 \left(1+(10^6{\rm K}/T )^3 \right) \frac{a/0.1 {\mu}{\rm m}}{n/{\rm cm^{-3}}} {\rm yr},
\label{timescale_d_destrcut}
\end{equation}
where $n$ and $T$ are the number density and temperature of the hot gas, respectively, and $a$ is the grain radius. The total stardust mass $M_{\rm D}(t)$ is reduced by thermal sputtering in the hot gas at the rate:

\begin{equation}
      \frac{{\rm d}M_{\rm D}(t)}{{\rm d}t} = -\frac{M_{D}{\rm (t)}}{{\tau_{\rm sput}}}.
      \label{dustsputtering}
\end{equation}

The temperature and density of the hot gas are derived from observations of extended X-ray emission. For simplicity, we assume single values for the electron density and temperature of the gas of $10^{-3}$ cm$^{-3}$ and $1.5\times 10^7$~K, respectively \citep{mathews03} resulting in $\tau_{\rm sput}=$100~Myr. Note that $\tau_{\rm sput}$ depends only weakly on temperature in the regime appropriate for the hot ISM of elliptical galaxies  and a value of $T=10^6$~K results in the time scale of 200~Myr. A similarly low value of the timescale of interstellar dust destruction, only 46~Myr, is derived for ETGs detected in FIR by {\it Spitzer} observations \citep{Clemens2010el}. For a comparison, we also ran calculations of the SSP evolution with a longer dust destruction timescale of 1~Gyr which corresponds to a lower gas density of $10^{-4}$ cm$^{-3}$. This long timescale may also account for the fact that many early type galaxies may harbour cold gas \citep{mathews03, alatalo13, young14}, where grains are protected for some time from the thermal sputtering and can survive longer. Another mechanism of dust destruction is inertial sputtering in SN shocks, which is thought to be the dominant mechanism of dust destruction in spiral galaxies. However, in a hot rarefied medium one SN destroys 20 times less dust compared to the local ISM conditions \citep{mckee89}. We therefore do not consider dust destruction by type Ia SNe and restrict our consideration to the thermal sputtering in hot gas. Dust mass in an early type galaxy can also be substantially reduced by the galactic winds (not considered in the present model). Our estimates should therefore be considered as the upper limit for the stardust mass.
 
 Fig.~\ref{dustmodel} compares the specific dust masses we have derived for the sample of \rg-E to the results of the SSP models \footnote{Value of the solar metallicity adopted here is Z$_{\odot}=0.014$ \citep{asplund09}.}. The data are grouped in three metallicity bins of [$0.5-1.5$] Z$_{\odot}$, [1.5-2.5] Z$_{\odot}$, and $> 2.5$ Z$_{\odot}$ and compared to three sets of SSP models with $Z=$ Z$_{\odot}$ (left panel), $Z=2 $Z $_{\odot}$ (middle panel), and $Z=3$ Z$_{\odot}$ (right panel). The specific dust masses of each \rg-E galaxy in the sample is plotted versus the mass weighted age of its stellar populations and the metallicity of each galaxy is obtained from the SDSS DR4 \citep{gallazi05}. The figure clearly shows that, as expected, SSP models with no dust destruction tend to over-predict the amount of dust in these ellipticals. On the other hand, more realistic models with dust sputtering fail to reproduce the observed $M_{\rm D}/M_{*}$ ratio even when a relatively long dust destruction timescale of $1$ Gyr is considered. The SSP models with dust destruction under-predicts the ratio of $M_{\rm D}/M_{*}$ by more than two order of magnitude. These estimates demonstrate that dust return into ISM from stellar sources is not sufficient to explain the observed $M_{\rm D}/M_{*}$. This implies an external origin of the dust via minor mergers and/or efficient dust growth in the dense ISM. 
 
The amount of dust in the sub-mm detected galaxies as well as its correlation with the present day star-formation rate (Fig.~\ref{fig:3panels}, panel b) suggests a connection between the dust and the dense ISM in agreement with \citet{alatalo13}, who find that the distribution of the CO and dust in nearby ETG is spatially correlated. The timescale for dust growth in molecular clouds is short and of the order of a few to several $10^{7}$ yrs  \citep{hirashita00}. We estimate an upper bound on the dust mass that may result from dust growth in the dense ISM in the following manner. Assuming a specific mass of molecular gas $M_{H_{2}}/M_{\star}$ of $0.01$ and a value of $0.06$ for the specific mass of the atomic gas $M_{HI}/M_{\star}$ (these are the observed upper limits in \cite{young14}), a dust-to-hydrogen mass ratio of $0.018$ (i.e., about 3 times the solar value), and a complete condensation of heavy elements into dust in the molecular gas, this yields a specific dust mass $M_{\rm D}/M_{\star}$ of $0.07\times0.018\approx 1.3\times10^{-3}$ which is only slightly higher than the largest specific dust masses measured for the sample of red Ellipticals that are displayed in Fig.~\ref{dustmodel}. This means that it is difficult, but not impossible, to explain the measured dust masses as resulting from grain growth in the dense gas inside the elliptical galaxies. It should be noted that dust growth does not preclude the role of minor mergers because the molecular gas may have an external origin \citep{davis11}.

\section{Conclusions}
\label{sec:conclusion}


In this work, we examine the properties of low redshift galaxies detected in 250\micron~ ($>$5$\sigma$) using H-ATLAS DR1 catalogue. 
We define two sub-samples of \rg and \bg galaxies based on \nuvr~ colours.  Our aim is to understand the nature of  the \rg subset in comparison to those in the \bg sub-sample.  We can summarize our findings as follow:

\begin{itemize}

\item Within the redshift range $0.01\leq z \leq 0.2$ of our sample, \rg sources with the UV-optical colour indices of \nuvr$\geq$4.5, constitute $\approx$4.2 per cent of the  total number of systems in H-ATLAS. The fraction of \rg sources increases with the galaxy stellar mass  such that in $\gtrsim$97 per cent of the \rg sample, M$_{*}\gtrsim$10$^{10}$M$_{\sun}$.\\

\item Following the visual inspection of galaxies, sources in the \rg sample were grouped into three categories of elliptical (E), spiral (S) and uncertain (U). We find that at least $\gtrsim$30 per cent of the \rg sources are of type E and more than $\gtrsim$40 per cent of sources belong to type S. \\

\item Both \bg and \rg sources, seem to occupy environments  with similar densities (e.g. having similar $\log(\Sigma_5)$ distributions) though in comparison to \bg and \rg objects of type S and U, a slightly larger fraction of \rg-E sources are in relatively denser regions with $\log(\Sigma_5/\rm{Mpc}^{-2})\gtrsim$1.5.\\

\item The SED analysis of galaxies in our sample based on MAGPHYS, reveals that the \rg galaxies (either type S or E) span a similar range of dust masses but different dust-to-stellar mass ratios in comparison to the \bg galaxies. The specific dust masses in the \bg and \rg-S galaxies are, on average, larger than those found for the \rg-E sample by a factor of 7$\times$ and  2$\times$ respectively. Similarly, galaxies of type E have lower levels of {\revised mean} SFR and SSFR in contrast to sources in the \bg and \rg-S samples. Furthermore, analysis of \fmu~ shows that unlike \bg galaxies where star-forming regions have the main contribution to the observed \submm fluxes, FIR emission in the \rg systems of type E is mainly from the dust in the ISM. \\

\item  The UV-optical colours of the \rg-⁠S sample could be the result of their highly inclined orientation and/⁠or a strong contribution of the old stellar population. However, in the current work we did not further investigate the contribution of each factor to the observed colour of the \rg-⁠S sources.\\

\item Finally, the comparison of specific dust masses ($M_{\rm D}/M_{*}$) of the \rg elliptical galaxies to the dust evolution in single stellar populations models excludes that the origin of the dust is from internal stellar sources. Dust growth in molecular clouds and/or gas and dust accretion through minor mergers provide more realistic and appealing alternatives \citep[e.g.,][]{gomez10,Smith12b}.\\

\end{itemize}
Our results show that there exist a population of early-type galaxies, containing a significant level of cold dust similar to those observed in blue/star-forming galaxies. The origin of dust in such early-type galaxies is likely to be of external origin (e.g. fuelled through mergers and tidal interactions). Hence, it is interesting to know  the difference between {\it red} galaxies which are detected in 250\micron~ and those without any \submm detection in the hope to find the mechanisms that are responsible for tuning the dust content in passive and/or early-type galaxies. 


\section*{Acknowledgments}
We would like to thank the anonymous referee for his valuable comments and suggestions 
which helped to improve this paper.

S. D. is supported by a Marie-Curie Intra European Fellowship under the European Community's Seventh Framework Program FP7/2007-2013 grant agreement no 627008. SH acknowledges support from the {\em Deutsche Forschungsgemeinschaft} (DFG) in the collaborative research project SFB881 “The Milky Way System” (subprojects B1).  RJI, LD and SJM acknowledge support from the European Research Council Advanced Grant, cosmicism. S. Z. acknowledge support by the Deutsche Forschungsgemeinschaft through SPP 1573: “Physics of the Interstellar Medium”. G. dZ. acknowledge financial support from ASI/INAF Agreement 2014-024-R.0 for the {\it Planck} LFI activity of Phase E2. K.~R. acknowledges support from the European Research Council Starting Grant SEDmorph (P.I. V.~Wild).

The {\it Herschel}-ATLAS is a project with {\it Herschel}; which is an ESA space observatory with science instruments provided by European-led Principal Investigator consortia and with important participation from NASA. The H-ATLAS website is http://www.h-atlas.org/. GAMA is a joint European-Australasian project based around a spectroscopic campaign using the Anglo-Australian Telescope. The GAMA input catalogue is based on data taken from the Sloan Digital Sky Survey and UKIRT Infrared Deep Sky Survey. Complementary imaging of the GAMA regions is being obtained by a number of independent survey programs including GALEX MIS, VST KIDS, VISTA VIKING, WISE, Herschel -ATLAS, GMRT and ASKAP providing UV to radio coverage. GAMA is funded by the STFC (UK), the ARC (Australia), the AAO, and the participating institutions. The GAMA website is http://www.gama-survey.org/. MAGPHYS is available via  http://www.iap.fr/magphys/magphys/MAGPHYS.html \\


\bibliographystyle{apacite}


\appendix
\section[]{SDSS postage-stamp images of \rg galaxies and their SED fits}
\label{appendix:gallery}

\begin{center}
\begin{table*}
\begin{tabular}{llllllllc}
index & HATLAS IAU ID & SDSS OBJID & SDSS Ra & SDSS Dec & \nuvr & $\log(\Sigma_5)$  & $i$ & type \\ 
 &  &  &  &  & [mag] & [Mpc$^{-2}$] & [deg] &  \\ \hline

1  &  HATLAS-J085450.2+021207  &  587727944563687568  &  8h54m50.22  &  +2h12m8.37  &  4.71  &  -0.693  &  56.3  &  U  \\ 
2  &  HATLAS-J092342.9+012056  &  587727942956220488  &  9h23m42.94  &  +1h20m57.21  &  5.06  &  0.099  &  38.67  &  S  \\ 
3  &  HATLAS-J084643.5+015034  &  587727944025964790  &  8h46m43.64  &  +1h50m35.95  &  5.44  &  0.997  &  70.88  &  S  \\ 
4  &  HATLAS-J084345.2-003205  &  588848899354329167  &  8h43m45.22  &  -0h32m4.59  &  5.28  &  -0.08  &  66.09  &  U  \\ 
5  &  HATLAS-J092110.3+021205  &  587726033304944826  &  9h21m10.43  &  +2h12m4.44  &  4.81  &  -1.143  &  82.92  &  S  \\ 
6  &  HATLAS-J084305.0+010858  &  587726032227008788  &  8h43m5.15  &  +1h8m55.59  &  4.66  &  0.055  &  57.29  &  S  \\ 
7  &  HATLAS-J092344.2-001113  &  588848899895591029  &  9h23m44.38  &  -0h11m14.06  &  4.72  &  -0.203  &  72.62  &  S  \\ 
8  &  HATLAS-J084139.5+015346  &  587726033300619494  &  8h41m39.55  &  +1h53m46.57  &  4.7  &  -0.484  &  34.33  &  U  \\ 
9  &  HATLAS-J084343.9-001243  &  587725074451595552  &  8h43m44.02  &  -0h12m43.98  &  4.67  &  0.035  &  73.62  &  S  \\ 
10  &  HATLAS-J085946.8-000019  &  588848899892969689  &  8h59m46.88  &  -0h0m20.2  &  4.75  &  -0.323  &  69.26  &  S  \\ 
11  &  HATLAS-J084713.9+012141  &  587727943489094075  &  8h47m14.09  &  +1h21m44.65  &  5.43  &  -0.648  &  35.45  &  E  \\ 
12  &  HATLAS-J090911.8+000030  &  587725074991218943  &  9h9m11.88  &  +0h0m28.79  &  5.16  &  -0.644  &  79.06  &  S  \\ 
13  &  HATLAS-J084632.0+001825  &  588848900428398906  &  8h46m32.24  &  +0h18m26.85  &  5.44  &  0.122  &  61.57  &  U  \\ 
14  &  HATLAS-J090952.3-003019  &  588848899357147464  &  9h9m52.4  &  -0h30m16.72  &  4.72  &  -1.013  &  48.67  &  E  \\ 
15  &  HATLAS-J085407.6+012716  &  587727943489880290  &  8h54m7.53  &  +1h27m18.01  &  4.52  &  -0.57  &  61.4  &  S  \\ 
16  &  HATLAS-J084625.7+014913  &  587727944025899418  &  8h46m25.84  &  +1h49m11.11  &  4.92  &  -0.427  &  53.09  &  U  \\ 
17  &  HATLAS-J083610.1+005604  &  587727942951043325  &  8h36m10.04  &  +0h56m0.54  &  4.72  &  0.665  &  53.86  &  U  \\ 
18  &  HATLAS-J091612.2-004200  &  587725073918263574  &  9h16m12.16  &  -0h41m58.08  &  4.8  &  -0.4  &  56.76  &  S  \\ 
19  &  HATLAS-J092158.0+023427  &  587727944566636774  &  9h21m58.05  &  +2h34m28.44  &  5.1  &  -1.051  &  39.17  &  E  \\ 
20  &  HATLAS-J084933.2+014340  &  587726032764600581  &  8h49m33.08  &  +1h43m40.89  &  4.78  &  -0.227  &  54.61  &  E  \\ 
21  &  HATLAS-J090752.4+012945  &  587727943491387551  &  9h7m52.23  &  +1h29m44.39  &  4.62  &  0.597  &  34.05  &  E  \\ 
22  &  HATLAS-J090929.3+020326  &  587727944028455086  &  9h9m29.56  &  +2h3m25.69  &  5.5  &  -0.356  &  62.22  &  U  \\ 
23  &  HATLAS-J084215.5+011605  &  587727943488569644  &  8h42m15.64  &  +1h16m5.77  &  4.67  &  0.221  &  74.86  &  S  \\ 
24  &  HATLAS-J084630.9+015620  &  587726033301143661  &  8h46m31.0  &  +1h56m21.44  &  4.63  &  -0.706  &  80.33  &  S  \\ 
25  &  HATLAS-J084324.4+005705  &  587727942951829819  &  8h43m24.52  &  +0h57m5.62  &  5.87  &  -0.245  &  37.25  &  E  \\ 
26  &  HATLAS-J085738.4+010741  &  587727942953402664  &  8h57m38.51  &  +1h7m41.34  &  5.02  &  0.702  &  68.5  &  S  \\ 
27  &  HATLAS-J091735.1+001931  &  588848900431741238  &  9h17m35.15  &  +0h19m30.52  &  5.06  &  -0.175  &  30.5  &  U  \\ 
28  &  HATLAS-J084929.1-005350  &  588010931369083190  &  8h49m29.3  &  -0h53m44.48  &  4.58  &  nan  &  38.67  &  U  \\ 
29  &  HATLAS-J085554.8-002832  &  588848899355639926  &  8h55m54.59  &  -0h28m26.59  &  6.41  &  0.669  &  46.49  &  E  \\ 
30  &  HATLAS-J091333.6-001508  &  587725074454806843  &  9h13m34.04  &  -0h15m9.56  &  4.74  &  -0.996  &  28.72  &  U  \\ 
31  &  HATLAS-J091143.6+012055  &  587726032230154446  &  9h11m43.76  &  +1h20m56.79  &  4.77  &  -1.367  &  60.52  &  U  \\ 
32  &  HATLAS-J092232.9-005813  &  587729151452774559  &  9h22m33.11  &  -0h58m13.64  &  5.03  &  2.057  &  29.99  &  E  \\ 
33  &  HATLAS-J085750.5-005517  &  587729151450022213  &  8h57m50.7  &  -0h55m17.26  &  4.97  &  0.825  &  68.92  &  S  \\ 
34  &  HATLAS-J084043.4+010814  &  587726032226746692  &  8h40m43.12  &  +1h8m11.83  &  4.78  &  0.632  &  27.14  &  U  \\ 
35  &  HATLAS-J092125.1-000341  &  588848899895328909  &  9h21m25.09  &  -0h3m43.62  &  4.86  &  -0.987  &  61.86  &  S  \\ 
36  &  HATLAS-J085311.5+005530  &  587727942952878410  &  8h53m11.59  &  +0h55m34.59  &  5.94  &  -0.282  &  70.27  &  S  \\ 
37  &  HATLAS-J085443.3+010539  &  587727942953074975  &  8h54m43.22  &  +1h5m45.35  &  5.07  &  -0.578  &  42.14  &  E  \\ 
38  &  HATLAS-J114923.8-010501  &  587748927628902552  &  11h49m23.54  &  -1h5m1.79  &  4.6  &  0.22  &  75.72  &  S  \\ 
39  &  HATLAS-J115841.9-011801  &  587724650867523744  &  11h58m41.95  &  -1h18m0.26  &  4.6  &  -0.833  &  81.55  &  S  \\ 
40  &  HATLAS-J121840.2-001522  &  587722982815891459  &  12h18m40.23  &  -0h15m23.27  &  4.64  &  -0.417  &  45.72  &  U  \\ 
41  &  HATLAS-J113955.6+013042  &  587728307494584346  &  11h39m55.86  &  +1h30m43.42  &  4.62  &  1.267  &  56.1  &  S  \\ 
42  &  HATLAS-J115256.8+012929  &  587728307495960699  &  11h52m57.0  &  +1h29m30.38  &  4.76  &  -0.244  &  71.53  &  S  \\ 
43  &  HATLAS-J120028.7-015138  &  587724650330849374  &  12h0m28.68  &  -1h51m38.87  &  5.21  &  -0.697  &  47.64  &  E  \\ 
44  &  HATLAS-J120844.2-003226  &  588848899376742632  &  12h8m44.22  &  -0h32m27.03  &  5.23  &  -0.717  &  53.36  &  U  \\ 
45  &  HATLAS-J120613.6-003423  &  588848899376480427  &  12h6m13.54  &  -0h34m23.79  &  4.54  &  -0.44  &  73.63  &  S  \\ 
46  &  HATLAS-J115448.1+000154  &  587748929240105086  &  11h54m48.05  &  +0h1m54.31  &  4.73  &  -0.281  &  58.09  &  E  \\ 
47  &  HATLAS-J121815.4-002151  &  587722982815826062  &  12h18m15.44  &  -0h21m53.46  &  4.63  &  0.002  &  62.11  &  S  \\ 
48  &  HATLAS-J121700.2-004455  &  587722982278824126  &  12h17m0.41  &  -0h44m57.05  &  4.78  &  0.602  &  75.37  &  S  \\ 
49  &  HATLAS-J115257.6+004210  &  588848900985651366  &  11h52m57.73  &  +0h42m9.72  &  5.17  &  -0.396  &  76.79  &  S  \\ 
50  &  HATLAS-J120028.9-000725  &  588848899912696073  &  12h0m28.87  &  -0h7m24.87  &  5.81  &  0.612  &  56.89  &  E  \\ 
51  &  HATLAS-J115754.8+001333  &  588848900449304761  &  11h57m54.83  &  +0h13m32.9  &  4.92  &  0.66  &  71.6  &  S  \\ 
52  &  HATLAS-J115442.0-005447  &  588848898838364283  &  11h54m42.05  &  -0h54m49.15  &  4.59  &  -0.51  &  58.0  &  U  \\ 
53  &  HATLAS-J114547.3-011709  &  587724650866081917  &  11h45m47.33  &  -1h17m8.13  &  4.57  &  0.401  &  65.02  &  S  \\ 
54  &  HATLAS-J115525.5-002039  &  587748928703299628  &  11h55m25.47  &  -0h20m42.63  &  4.6  &  0.205  &  60.34  &  S  \\ 
55  &  HATLAS-J114837.1-011246  &  587748927628837012  &  11h48m37.19  &  -1h12m46.2  &  6.28  &  1.702  &  39.33  &  E  \\ 
56  &  HATLAS-J115827.6+004304  &  588848900986241088  &  11h58m27.7  &  +0h43m4.46  &  6.08  &  -0.541  &  56.33  &  E  \\ 
57  &  HATLAS-J121636.4-005723  &  588848898840723542  &  12h16m36.51  &  -0h57m21.43  &  5.19  &  -0.79  &  63.45  &  E  \\ 
58  &  HATLAS-J115122.7+000702  &  587748929239711890  &  11h51m22.64  &  +0h7m2.43  &  4.68  &  -0.597  &  23.1  &  U  \\ 
59  &  HATLAS-J121747.1+003553  &  587722983889502322  &  12h17m47.17  &  +0h35m51.09  &  4.86  &  -0.583  &  72.32  &  S  \\ 
60  &  HATLAS-J120454.4+011402  &  588848901523832979  &  12h4m54.65  &  +1h14m2.7  &  5.35  &  -0.172  &  26.47  &  E  \\

\end{tabular}
\caption{List of all \rg galaxies detected in HATLAS.}
  \label{table:gallery-01}
  \end{table*}
\end{center}

\begin{center}
\begin{table*}
\begin{tabular}{llllllllc}
index & HATLAS IAU ID & SDSS OBJID & SDSS Ra & SDSS Dec & \nuvr & $\log(\Sigma_5)$  & $i$ & type \\ 
 &  &  &  &  & [mag] & [Mpc$^{-2}$] & [deg] &  \\ \hline
 
61  &  HATLAS-J114750.4-013710  &  587725041701159100  &  11h47m50.38  &  -1h37m11.31  &  4.86  &  0.558  &  49.64  &  U  \\ 
62  &  HATLAS-J114828.1+001825  &  588848900448256260  &  11h48m28.25  &  +0h18m22.94  &  4.7  &  nan  &  56.22  &  E  \\ 
63  &  HATLAS-J120212.5-014032  &  587724650331045959  &  12h2m12.24  &  -1h40m31.17  &  4.75  &  -0.764  &  63.17  &  S  \\ 
64  &  HATLAS-J114930.0-010511  &  587748927628902442  &  11h49m30.15  &  -1h5m11.46  &  5.58  &  0.277  &  39.47  &  E  \\ 
65  &  HATLAS-J115053.9-010830  &  587722981739069591  &  11h50m53.76  &  -1h8m29.65  &  4.93  &  0.115  &  37.35  &  U  \\ 
66  &  HATLAS-J120008.3-003950  &  587748928166953080  &  12h0m8.17  &  -0h39m48.21  &  4.94  &  0.066  &  60.54  &  U  \\ 
67  &  HATLAS-J120048.1-011117  &  587748927630147744  &  12h0m48.28  &  -1h11m17.6  &  5.01  &  -0.461  &  47.56  &  E  \\ 
68  &  HATLAS-J113836.4-013713  &  587724650328424633  &  11h38m36.27  &  -1h37m14.05  &  4.52  &  0.554  &  22.58  &  E  \\ 
69  &  HATLAS-J122026.8-011046  &  587722981742280865  &  12h20m26.87  &  -1h10m47.28  &  4.67  &  0.446  &  34.85  &  U  \\ 
70  &  HATLAS-J120844.8+001220  &  587748929241612470  &  12h8m44.83  &  +0h12m21.46  &  4.98  &  -0.572  &  42.44  &  U  \\ 
71  &  HATLAS-J121001.7-011516  &  587724650868768886  &  12h10m1.61  &  -1h15m17.01  &  5.68  &  -0.833  &  58.76  &  S  \\ 
72  &  HATLAS-J113919.1-012012  &  587724650865361032  &  11h39m18.95  &  -1h20m18.19  &  5.05  &  -0.521  &  55.2  &  U  \\ 
73  &  HATLAS-J114318.5-004414  &  587748928165118125  &  11h43m18.61  &  -0h44m17.11  &  4.53  &  -0.539  &  51.33  &  U  \\ 
74  &  HATLAS-J120140.5+005138  &  587748930314567848  &  12h1m40.15  &  +0h51m38.71  &  5.01  &  -0.644  &  61.67  &  U  \\ 
75  &  HATLAS-J121823.6-013038  &  587725041704501421  &  12h18m23.51  &  -1h30m37.86  &  4.83  &  -0.167  &  59.44  &  U  \\ 
76  &  HATLAS-J120535.5+010445  &  588848901523898501  &  12h5m35.33  &  +1h4m44.34  &  5.53  &  0.479  &  49.35  &  U  \\ 
77  &  HATLAS-J114526.8-002708  &  588848899374186712  &  11h45m26.58  &  -0h27m11.56  &  5.32  &  -0.914  &  29.57  &  E  \\ 
78  &  HATLAS-J114849.6-005941  &  588848898837708980  &  11h48m49.57  &  -0h59m40.53  &  4.88  &  -0.6  &  53.97  &  U  \\ 
79  &  HATLAS-J114609.3-010205  &  588848898837446812  &  11h46m9.18  &  -1h2m6.83  &  4.88  &  0.585  &  63.8  &  S  \\ 
80  &  HATLAS-J120246.1+002207  &  588848900449829017  &  12h2m46.51  &  +0h22m3.61  &  6.64  &  -0.207  &  53.62  &  S  \\ 
81  &  HATLAS-J120406.6+001411  &  588848900449960274  &  12h4m6.52  &  +0h14m9.77  &  4.98  &  -0.117  &  72.22  &  S  \\ 
82  &  HATLAS-J145112.4-002724  &  588848899394568318  &  14h51m12.4  &  -0h27m24.76  &  4.71  &  0.187  &  75.01  &  S  \\ 
83  &  HATLAS-J143224.5+005041  &  587722984441118986  &  14h32m24.62  &  +0h50m41.14  &  4.9  &  -0.133  &  86.81  &  S  \\ 
84  &  HATLAS-J141501.6-005136  &  588848898853699826  &  14h15m1.74  &  -0h51m36.46  &  5.33  &  -0.412  &  82.77  &  S  \\ 
85  &  HATLAS-J143143.3-011418  &  587729972324073647  &  14h31m43.38  &  -1h14m19.78  &  4.84  &  -1.137  &  77.59  &  S  \\ 
86  &  HATLAS-J143801.4-001217  &  588848899929997456  &  14h38m1.53  &  -0h12m18.13  &  4.65  &  -0.479  &  69.75  &  S  \\ 
87  &  HATLAS-J141126.2+011711  &  587726014009573415  &  14h11m26.23  &  +1h17m11.47  &  5.55  &  0.777  &  19.37  &  E  \\ 
88  &  HATLAS-J142004.5-001852  &  587722982829130030  &  14h20m4.67  &  -0h18m53.29  &  4.6  &  0.053  &  33.78  &  U  \\ 
89  &  HATLAS-J141611.6+015204  &  587726032263446738  &  14h16m11.83  &  +1h52m4.72  &  5.5  &  -0.575  &  62.73  &  U  \\ 
90  &  HATLAS-J143012.5+001400  &  588848900465951018  &  14h30m12.5  &  +0h14m2.81  &  4.81  &  0.855  &  58.65  &  S  \\ 
91  &  HATLAS-J144810.4+012203  &  587726014550442257  &  14h48m10.5  &  +1h22m1.93  &  4.57  &  -0.393  &  68.64  &  S  \\ 
92  &  HATLAS-J141446.6-000417  &  588848899927441586  &  14h14m46.6  &  -0h4m17.37  &  5.26  &  -0.764  &  59.98  &  S  \\ 
93  &  HATLAS-J142926.0+012315  &  587726031728017631  &  14h29m26.06  &  +1h23m16.62  &  4.56  &  -0.658  &  57.74  &  S  \\ 
94  &  HATLAS-J141727.9+002857  &  587722983902609591  &  14h17m27.97  &  +0h28m57.99  &  5.19  &  0.713  &  40.76  &  E  \\ 
95  &  HATLAS-J141310.5+014618  &  587726014546641064  &  14h13m10.5  &  +1h46m17.11  &  5.57  &  2.006  &  44.1  &  E  \\ 
96  &  HATLAS-J144224.0+005430  &  587722984442232848  &  14h42m23.61  &  +0h54m28.79  &  5.01  &  -0.433  &  41.32  &  E  \\ 
97  &  HATLAS-J142113.4-002756  &  588848899391226106  &  14h21m13.45  &  -0h27m59.63  &  4.94  &  -0.479  &  32.78  &  E  \\ 
98  &  HATLAS-J142015.8+010252  &  587722984439808094  &  14h20m15.91  &  +1h2m51.5  &  4.81  &  0.17  &  65.57  &  S  \\ 
99  &  HATLAS-J141539.0-002649  &  588848899390636315  &  14h15m39.07  &  -0h26m51.7  &  4.85  &  -0.098  &  57.74  &  U  \\ 
100  &  HATLAS-J142429.3+015829  &  587726015084757174  &  14h24m29.34  &  +1h58m31.01  &  4.82  &  0.175  &  74.27  &  S  \\ 
101  &  HATLAS-J142856.4+002130  &  588848900465819923  &  14h28m56.56  &  +0h21m32.39  &  5.67  &  -0.635  &  25.6  &  E  \\ 
102  &  HATLAS-J142613.8-011122  &  587729972323483911  &  14h26m13.74  &  -1h11m24.01  &  5.29  &  0.195  &  39.73  &  E  \\ 
103  &  HATLAS-J143052.0+011836  &  587726031728214195  &  14h30m52.04  &  +1h18m34.61  &  4.97  &  -0.672  &  71.24  &  S  \\ 
104  &  HATLAS-J143731.7+000341  &  587722983367901556  &  14h37m31.92  &  +0h3m39.01  &  4.63  &  -0.87  &  72.71  &  S  \\ 
105  &  HATLAS-J144532.2-010921  &  587729972325646543  &  14h45m32.17  &  -1h9m20.9  &  4.75  &  -0.757  &  79.16  &  S  \\ 
106  &  HATLAS-J144346.1+004306  &  588848901004329189  &  14h43m46.24  &  +0h43m4.43  &  4.59  &  -0.767  &  61.13  &  U  \\ 
107  &  HATLAS-J140753.5-001931  &  587722982827819184  &  14h7m53.34  &  -0h19m27.74  &  4.5  &  -0.396  &  27.39  &  E  \\ 
108  &  HATLAS-J142831.0+014541  &  587726032264822925  &  14h28m31.19  &  +1h45m40.78  &  5.53  &  -0.599  &  35.48  &  E  \\ 
109  &  HATLAS-J144718.4-010621  &  587729972325843159  &  14h47m18.4  &  -1h6m18.83  &  4.63  &  0.055  &  48.6  &  E  \\ 
110  &  HATLAS-J142517.4-010304  &  587722981755977936  &  14h25m17.4  &  -1h3m6.24  &  5.18  &  0.139  &  33.86  &  U  \\ 
111  &  HATLAS-J142437.5-013819  &  587729971786481829  &  14h24m37.35  &  -1h38m20.15  &  5.4  &  0.688  &  31.13  &  E  \\ 
112  &  HATLAS-J145123.6+000025  &  587722983369474066  &  14h51m23.42  &  +0h0m25.48  &  4.94  &  -0.627  &  44.69  &  E  \\ 
113  &  HATLAS-J141353.0-004527  &  587722982291603595  &  14h13m53.48  &  -0h45m27.18  &  5.07  &  0.484  &  22.97  &  E  \\ 
114  &  HATLAS-J141325.9-004923  &  587722982291538161  &  14h13m25.85  &  -0h49m23.89  &  5.12  &  0.36  &  43.99  &  E  \\ 
115  &  HATLAS-J145216.9+010631  &  587726014014030018  &  14h52m16.66  &  +1h6m34.3  &  5.01  &  -0.721  &  32.7  &  E  \\ 
116  &  HATLAS-J142512.3-001858  &  587722982829719819  &  14h25m12.49  &  -0h19m0.67  &  4.92  &  -0.287  &  47.84  &  E  \\ 
117  &  HATLAS-J141516.7-003941  &  587722982291734808  &  14h15m16.49  &  -0h39m40.61  &  5.29  &  -0.088  &  70.96  &  S  \\

\end{tabular}
\caption{Table.~\ref{table:gallery-01} Continued.}
  \label{table:gallery-02}
  \end{table*}
\end{center}

\begin{center}
\begin{table*}
\begin{tabular}{llllllll}

index & HATLAS IAU ID & $\log$(\mstar) & $\log$(SFR) & $\log$(SFR/$\rm{M_*}$) & $\log$(\mdust) & $\log$(\Mdust) & f$_{\mu}$ \\
 &  &  & [M$_{\odot}$yr $^{-1}$] & [yr $^{-1}$] &  &  &  \\ \hline
1  &  HATLAS-J114923.8-010501  &  10.92  &  -0.09  &  -11.01  &  7.51  &  -3.41  &  0.72  \\ 
2  &  HATLAS-J115841.9-011801  &  10.98  &  0.32  &  -10.66  &  8.08  &  -2.89  &  0.78  \\ 
3  &  HATLAS-J121840.2-001522  &  11.29  &  -0.21  &  -11.5  &  7.77  &  -3.52  &  0.78  \\ 
4  &  HATLAS-J113955.6+013042  &  11.2  &  -0.04  &  -11.25  &  7.82  &  -3.38  &  0.67  \\ 
5  &  HATLAS-J115256.8+012929  &  10.61  &  -0.25  &  -10.86  &  7.97  &  -2.64  &  0.78  \\ 
6  &  HATLAS-J120028.7-015138  &  10.95  &  -0.79  &  -11.75  &  7.87  &  -3.08  &  0.83  \\ 
7  &  HATLAS-J120844.2-003226  &  11.05  &  -1.04  &  -12.08  &  7.39  &  -3.65  &  0.82  \\ 
8  &  HATLAS-J120613.6-003423  &  10.59  &  0.02  &  -10.57  &  7.35  &  -3.24  &  0.68  \\ 
9  &  HATLAS-J115448.1+000154  &  11.02  &  -0.35  &  -11.37  &  8.11  &  -2.9  &  0.78  \\ 
10  &  HATLAS-J121815.4-002151  &  10.51  &  -2.67  &  -13.18  &  7.32  &  -3.18  &  0.98  \\ 
11  &  HATLAS-J121700.2-004455  &  10.68  &  -0.84  &  -11.52  &  7.86  &  -2.83  &  0.74  \\ 
12  &  HATLAS-J115257.6+004210  &  10.85  &  -0.62  &  -11.47  &  7.82  &  -3.03  &  0.8  \\ 
13  &  HATLAS-J120028.9-000725  &  11.35  &  -0.87  &  -12.22  &  7.86  &  -3.49  &  0.8  \\ 
14  &  HATLAS-J115754.8+001333  &  10.55  &  -0.34  &  -10.89  &  7.44  &  -3.11  &  0.86  \\ 
15  &  HATLAS-J115442.0-005447  &  10.66  &  -2.27  &  -12.93  &  8.28  &  -2.37  &  0.85  \\ 
16  &  HATLAS-J114547.3-011709  &  10.68  &  -0.23  &  -10.91  &  8.05  &  -2.63  &  0.72  \\ 
17  &  HATLAS-J115525.5-002039  &  11.11  &  0.2  &  -10.91  &  7.67  &  -3.44  &  0.74  \\ 
18  &  HATLAS-J114837.1-011246  &  11.52  &  -1.37  &  -12.89  &  8.26  &  -3.27  &  0.86  \\ 
19  &  HATLAS-J115827.6+004304  &  10.92  &  -2.85  &  -13.77  &  6.65  &  -4.26  &  0.85  \\ 
20  &  HATLAS-J121636.4-005723  &  10.9  &  -1.41  &  -12.31  &  7.1  &  -3.8  &  0.84  \\ 
21  &  HATLAS-J115122.7+000702  &  10.88  &  -0.63  &  -11.51  &  6.87  &  -4.01  &  0.64  \\ 
22  &  HATLAS-J121747.1+003553  &  10.71  &  -0.46  &  -11.17  &  7.3  &  -3.41  &  0.76  \\ 
23  &  HATLAS-J120454.4+011402  &  11.02  &  -1.03  &  -12.05  &  7.41  &  -3.61  &  0.8  \\ 
24  &  HATLAS-J114750.4-013710  &  10.71  &  -0.95  &  -11.66  &  7.16  &  -3.55  &  0.94  \\ 
25  &  HATLAS-J114828.1+001825  &  11.14  &  -0.16  &  -11.3  &  7.78  &  -3.36  &  0.99  \\ 
26  &  HATLAS-J120212.5-014032  &  10.95  &  -0.54  &  -11.49  &  8.07  &  -2.88  &  0.83  \\ 
27  &  HATLAS-J114930.0-010511  &  10.81  &  -1.89  &  -12.69  &  6.46  &  -4.35  &  0.82  \\ 
28  &  HATLAS-J115053.9-010830  &  10.73  &  -0.64  &  -11.37  &  8.2  &  -2.53  &  0.75  \\ 
29  &  HATLAS-J120008.3-003950  &  9.6  &  -1.77  &  -11.36  &  5.91  &  -3.69  &  0.16  \\ 
30  &  HATLAS-J120048.1-011117  &  10.83  &  -0.44  &  -11.27  &  8.06  &  -2.77  &  0.69  \\ 
31  &  HATLAS-J113836.4-013713  &  10.89  &  -0.59  &  -11.48  &  6.87  &  -4.02  &  0.68  \\ 
32  &  HATLAS-J122026.8-011046  &  10.68  &  -1.45  &  -12.13  &  7.56  &  -3.13  &  0.85  \\ 
33  &  HATLAS-J121001.7-011516  &  10.29  &  -2.43  &  -12.72  &  7.32  &  -2.97  &  0.85  \\ 
34  &  HATLAS-J113919.1-012012  &  10.5  &  -1.19  &  -11.69  &  7.36  &  -3.15  &  0.84  \\ 
35  &  HATLAS-J114318.5-004414  &  10.54  &  -0.69  &  -11.23  &  7.21  &  -3.34  &  0.71  \\ 
36  &  HATLAS-J120140.5+005138  &  10.94  &  -0.65  &  -11.58  &  7.41  &  -3.53  &  0.91  \\ 
37  &  HATLAS-J121823.6-013038  &  10.57  &  -1.14  &  -11.72  &  6.95  &  -3.63  &  0.84  \\ 
38  &  HATLAS-J120535.5+010445  &  10.75  &  -1.54  &  -12.29  &  7.36  &  -3.39  &  0.99  \\ 
39  &  HATLAS-J114526.8-002708  &  10.99  &  -1.13  &  -12.11  &  7.2  &  -3.78  &  0.86  \\ 
40  &  HATLAS-J114849.6-005941  &  11.25  &  -0.43  &  -11.68  &  7.54  &  -3.71  &  1.0  \\ 
41  &  HATLAS-J114609.3-010205  &  10.66  &  -0.89  &  -11.55  &  7.57  &  -3.1  &  0.87  \\ 
42  &  HATLAS-J120246.1+002207  &  11.27  &  -1.42  &  -12.69  &  7.66  &  -3.61  &  0.81  \\ 
43  &  HATLAS-J120406.6+001411  &  10.92  &  -0.61  &  -11.53  &  6.93  &  -3.99  &  0.81  \\ 
44  &  HATLAS-J145112.4-002724  &  10.63  &  -0.44  &  -11.07  &  7.68  &  -2.95  &  0.75  \\ 
45  &  HATLAS-J143224.5+005041  &  11.29  &  0.03  &  -11.26  &  8.38  &  -2.91  &  0.79  \\ 
46  &  HATLAS-J141501.6-005136  &  10.61  &  -0.79  &  -11.4  &  7.37  &  -3.24  &  0.81  \\ 
47  &  HATLAS-J143143.3-011418  &  10.6  &  -0.43  &  -11.02  &  6.98  &  -3.61  &  0.73  \\ 
48  &  HATLAS-J143801.4-001217  &  10.95  &  0.02  &  -10.93  &  8.11  &  -2.84  &  0.79  \\ 
49  &  HATLAS-J141126.2+011711  &  11.03  &  -1.51  &  -12.54  &  6.87  &  -4.16  &  0.88  \\ 
50  &  HATLAS-J142004.5-001852  &  10.36  &  -0.21  &  -10.57  &  7.05  &  -3.31  &  0.71  \\ 
51  &  HATLAS-J141611.6+015204  &  11.03  &  -1.88  &  -12.91  &  8.12  &  -2.91  &  0.85  \\ 
52  &  HATLAS-J143012.5+001400  &  10.69  &  -0.69  &  -11.38  &  7.37  &  -3.32  &  0.76  \\ 
53  &  HATLAS-J144810.4+012203  &  9.89  &  -0.96  &  -10.85  &  6.56  &  -3.33  &  0.72  \\ 
54  &  HATLAS-J142926.0+012315  &  11.09  &  -0.17  &  -11.27  &  7.88  &  -3.21  &  0.71  \\ 
55  &  HATLAS-J141727.9+002857  &  10.97  &  -1.35  &  -12.31  &  7.15  &  -3.81  &  0.96  \\ 
56  &  HATLAS-J141310.5+014618  &  11.03  &  -0.83  &  -11.86  &  6.96  &  -4.06  &  0.72  \\ 
57  &  HATLAS-J144224.0+005430  &  10.95  &  -0.84  &  -11.79  &  6.85  &  -4.1  &  0.82  \\ 
58  &  HATLAS-J142113.4-002756  &  11.09  &  -0.29  &  -11.38  &  7.53  &  -3.56  &  0.83  \\ 
59  &  HATLAS-J142015.8+010252  &  10.9  &  -0.06  &  -10.96  &  7.52  &  -3.38  &  0.77  \\ 
60  &  HATLAS-J141539.0-002649  &  10.73  &  -0.44  &  -11.17  &  7.32  &  -3.41  &  0.78  \\ 
\end{tabular}
\caption{MAGPHYS output parameters for the \rg galaxies having {\it WISE} observed photometric data.}
  \label{table:sed-params-01}
  \end{table*}
\end{center}

\begin{center}
\begin{table*}
\begin{tabular}{llllllll}

index & HATLAS IAU ID & $\log$(\mstar) & $\log$(SFR) & $\log$(SFR/$\rm{M_*}$) & $\log$(\mdust) & $\log$(\Mdust) & f$_{\mu}$ \\
 &  &  & [M$_{\odot}$yr $^{-1}$] & [yr $^{-1}$] &  &  &  \\ \hline
61  &  HATLAS-J142429.3+015829  &  10.19  &  -0.75  &  -10.94  &  6.78  &  -3.4  &  0.76  \\ 
62  &  HATLAS-J142856.4+002130  &  11.45  &  -0.39  &  -11.83  &  7.88  &  -3.57  &  0.78  \\ 
63  &  HATLAS-J142613.8-011122  &  10.87  &  -0.78  &  -11.65  &  6.93  &  -3.95  &  0.8  \\ 
64  &  HATLAS-J143052.0+011836  &  11.19  &  -0.05  &  -11.24  &  8.01  &  -3.18  &  0.74  \\ 
65  &  HATLAS-J143731.7+000341  &  10.59  &  -0.43  &  -11.02  &  7.19  &  -3.4  &  0.73  \\ 
66  &  HATLAS-J144532.2-010921  &  10.38  &  -1.19  &  -11.57  &  7.32  &  -3.05  &  0.96  \\ 
67  &  HATLAS-J144346.1+004306  &  10.25  &  -0.65  &  -10.89  &  6.77  &  -3.47  &  0.64  \\ 
68  &  HATLAS-J140753.5-001931  &  10.91  &  -0.64  &  -11.55  &  6.92  &  -3.98  &  0.75  \\ 
69  &  HATLAS-J142831.0+014541  &  11.06  &  -1.06  &  -12.11  &  7.5  &  -3.56  &  0.88  \\ 
70  &  HATLAS-J144718.4-010621  &  11.2  &  -0.08  &  -11.28  &  7.51  &  -3.68  &  0.88  \\ 
71  &  HATLAS-J142517.4-010304  &  11.06  &  -0.55  &  -11.61  &  7.65  &  -3.41  &  0.84  \\ 
72  &  HATLAS-J142437.5-013819  &  11.0  &  -0.86  &  -11.86  &  7.28  &  -3.72  &  0.76  \\ 
73  &  HATLAS-J145123.6+000025  &  10.1  &  -1.39  &  -11.49  &  6.04  &  -4.06  &  0.69  \\ 
74  &  HATLAS-J141353.0-004527  &  11.17  &  -0.64  &  -11.81  &  7.86  &  -3.31  &  0.85  \\ 
75  &  HATLAS-J141325.9-004923  &  10.87  &  -0.73  &  -11.6  &  7.94  &  -2.93  &  0.78  \\ 
76  &  HATLAS-J145216.9+010631  &  11.06  &  -0.4  &  -11.46  &  7.2  &  -3.86  &  0.77  \\ 
77  &  HATLAS-J142512.3-001858  &  10.95  &  -1.16  &  -12.11  &  7.65  &  -3.3  &  0.86  \\ 
78  &  HATLAS-J141516.7-003941  &  10.93  &  -1.44  &  -12.37  &  7.65  &  -3.28  &  1.0  \\ 
\end{tabular}
\caption{Table.~\ref{table:sed-params-01} Continued.}
  \label{table:sed-params-02}
  \end{table*}
\end{center}

\end{document}